\documentclass[amsmath,
 amssymb,
 aps,
 nofootinbib,
 preprintnumbers
]{revtex4-1}
\usepackage[utf8]{inputenc}
\usepackage{simplewick}
\usepackage{footnote}
\usepackage{graphicx}
\usepackage{dcolumn}
\usepackage{bm}
\usepackage{stmaryrd}
\usepackage{amsmath}
\usepackage{amssymb}
\usepackage{mathtools}
\usepackage{bbm}
\usepackage{amsfonts}
\usepackage{xcolor}
\usepackage{footmisc}
\usepackage{comment}
\usepackage{ulem}
\usepackage{slashed}
\usepackage{multirow}
\usepackage{appendix}
\usepackage{float}
\usepackage{braket}
\usepackage[linktocpage=true,bookmarksnumbered=true,colorlinks=true,plainpages,linkcolor=blue,citecolor=red]{hyperref}
\usepackage{cancel}

\begin{document}
\preprint{ADP-23-26/T1235, MSUHEP-23-029}
\title{Scattering Amplitudes of Massive Spin-2 Kaluza-Klein States with Matter}

\author{R. Sekhar Chivukula$^{a}$}
\author{Joshua A. Gill$^{b}$}
\author{Kirtimaan A. Mohan$^{c}$}
\author{Dipan~Sengupta$^{b}$}
\author{Elizabeth H. Simmons$^{a}$}
\author{Xing Wang$^{a}$}

\affiliation{$^{a}$Department of Physics and Astronomy, University of California, San Diego, 9500 Gilman Drive, La Jolla, CA-92093, USA
}
\affiliation{$^{b}$ARC Centre of Excellence for Dark Matter Particle Physics, Department of Physics, University of Adelaide, South Australia 5005, Australia}
\affiliation{$^{c}$Department of Physics and Astronomy, 
Michigan State University\\
567 Wilson Road, East Lansing, MI-48824, USA
}

\begin{abstract}

We perform a comprehensive analysis of the scattering of matter and gravitational Kaluza-Klein (KK) modes in five-dimensional gravity theories. We consider matter localized on a brane as well as in the bulk of the extra dimension for scalars, fermions and vectors respectively, and consider an arbitrary warped background. While naive power-counting suggests that there are amplitudes which grow as fast as ${\cal O}(s^3)$ [where $s$ is the center-of-mass scattering energy-squared], we demonstrate that cancellations between the various contributions result in a total amplitude which grows no faster than ${\cal O}(s)$. Extending previous work on the self-interactions of the gravitational KK modes, we show that these cancellations occur due to sum-rule relations between the couplings and the masses of the modes that can be proven from the properties of the mode equations describing the gravity and matter wavefunctions. We demonstrate that these properties are tied to  the underlying diffeomorphism invariance of the five-dimensional theory. We discuss how our results generalize when the size of the extra dimension is stabilized via the Goldberger-Wise mechanism. Our conclusions are of particular relevance for freeze-out and freeze-in relic abundance calculations for dark matter models including a spin-2 portal arising from an underlying five-dimensional theory.
\end{abstract}

\maketitle

\tableofcontents

\pagebreak

\section{Introduction}
In recent years there has been a revival of interest in the phenomenology and cosmology of models with compactified extra dimensions: Kaluza-Klein(KK) theories \cite{Kaluza:1921tu}. The revival of KK theories was motivated by new solutions to the hierarchy problem which relate the scales associated with gravity and electroweak symmetry breaking. 
These included models with flat (``large") extra dimensions \cite{Antoniadis:1990ew,ArkaniHamed:1998rs}, as well as those with a ``small" warped extra dimension based on a slice of Anti-de-Sitter (AdS) space, known as Randall-Sundrum (RS) models \cite{Randall:1999ee,Randall:1999vf}. Extra dimensions have been used to address the flavor puzzle (see, for example, \cite{Arkani-Hamed:1999pwe,Agashe:2004cp}) to provide a path toward understanding the electroweak phase transition \cite{Creminelli:2001th,Nardini:2007me}, and to provide candidates for a dark sector. (For reviews of these developments see  \cite{Rattazzi:2003ea,Csaki:2004ay,Gabadadze:2003ii,Quevedo:2010ui}.) More recently, motivated specifically by dark matter and other cosmological considerations, new beyond the standard-model (BSM) scenarios have emerged in which extra dimensions play a crucial role, ranging from those including dark matter freeze-out \cite{deGiorgi:2021xvm,Folgado:2019sgz} and freeze-in \cite{Bernal:2020fvw, Bernal:2020yqg, deGiorgi:2022yha}, to continuum dark matter \cite{Csaki:2021gfm}, the holographic axion \cite{Cox:2019rro}, and dark dimensions in the Swampland conjecture \cite{Gonzalo:2022jac}. \looseness=-1 

\begin{figure}
\includegraphics[]{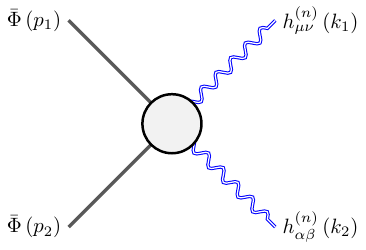}
\includegraphics[]{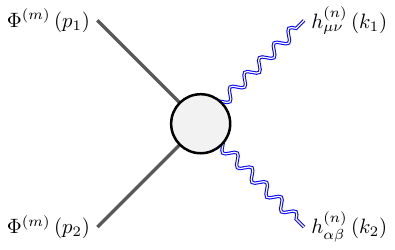}
\caption{An example of $2\to 2$ scattering of matter particles (where $\bar{\Phi} = \bar{S}, \chi, \bar{V}$ and $\Phi = S, \psi, V$) on the brane (left) and in the bulk (right) to spin-2 KK modes. The circle in the middle indicates all intermediate states, and $s,t,u$ and contact diagrams.}
\label{fig:GP}
\end{figure}

In many BSM scenarios a key ingredient is the calculation of squared matrix elements for the scattering of matter (including possible KK excitations) with massive spin-2 Kaluza-Klein graviton states. In particular, these scattering amplitudes  are of specific relevance for freeze-out and freeze-in relic abundance calculations for dark matter models including spin-2 portals, as well as for the study of the potential collider signatures of such theories. Calculations involving massive spin-2 states, however, are plagued by (as we show, potentially anomalous) contributions that grow rapidly with the center-of-mass energy of the scattering process. For example, calculations that involve the production of massive spin-2 KK particles in the final state from matter particles, such as the ones shown in Figure \ref{fig:GP}, have contributions due to the helicity-0 mode of the massive spin-2 states
that naively grow like $s^{3}/M_{\rm KK}^{4}$, where $s$ is the center of mass energy-squared of the scattering process and $M_{\rm KK}$ the mass of the spin-2 KK modes. This anomalous high energy behaviour - note the anomalous dependence on the low-energy scale $M_{\rm KK}$ - has been used to estimate observables like the relic density as well as direct detection rates for spin-2 KK mediated dark matter models \cite{Garny:2015sjg,Lee:2013bua,Folgado:2019sgz}. \looseness=-1

As we show in this paper, while it is true that the contributions from {\it individual diagrams} to the scattering of matter with massive spin-2 states can indeed grow as fast as ${\cal O}(s^3)$, a complete analysis using the underlying gravitational theory uncovers a cancellation between different contributions so that the {\it full amplitude} grows only like ${\cal O}(s)$. Therefore, phenomenological results based on the naive dimensional analyses of the individual contributions to the scattering amplitude \cite{Garny:2015sjg,Lee:2013bua,Folgado:2019sgz} lead to erroneous conclusions.\looseness=-1

This work is an extension of previous analyses \cite{SekharChivukula:2019qih,SekharChivukula:2019yul,Chivukula:2020hvi,Chivukula:2022kju} conducted by the authors and their collaborators on the properties of the amplitudes for the scattering of massive spin-2 states among themselves. In Kaluza-Klein theories we have shown  that the  scattering amplitudes involving spin-2 KK mode self-interactions grow only like ${\cal O}(s)$ despite there being individual contributions that grow as fast as ${\cal O}(s^5)$. We showed that the full amplitudes grow as $s/M_{\rm Pl}^{2}$
for flat extra dimensions (toroidal compactification) with $M_{\rm Pl}$ being the 4 dimensional Planck mass, and as
$s/\Lambda_{\pi}^{2}$ for RS compactification,  with  $\Lambda_{\pi}$ being the effective scale\footnote{$\Lambda_{\pi}=M_{\rm Pl}e^{-k r_{c}\pi}$, where $k$ is the curvature and $r_{c}$ is the radius of curvature.} of the compactified Randall-Sundrum model. 

In this work we extend our previous analyses to compute matter interactions with the gravitational sector in extra dimensions, show that the anomalous high-energy growth cancels, and show that the physical amplitudes grow only as fast as ${\cal O}(s)$. We perform a comprehensive analysis of the scattering of matter and gravitational modes in extra-dimensional theories: we consider matter localized on the brane as well as in the bulk of the extra dimensions for scalars, fermions and vectors respectively, and consider an arbitrary warped background (in which case flat or toroidal compactification is a special case where the curvature goes to zero). We show that while individual $2 \to 2$ scattering diagrams ($s,t,u$ and contact, see Figure, \ref{fig:branescalar} for example) grow anomalously, delicate cancellations enforced by a series of sum rules ensure that the overall amplitude is well behaved.  A special case of the computations reported here has been performed\footnote{After this work had been submitted and announced on arXiv, we were informed of \cite{deGiorgi:2023mdy}. Aside from the computation of the production of brane scalar particles from KK gaviton annhilation previously published in \cite{deGiorgi:2020qlg} and cited here, Ref.~\cite{deGiorgi:2023mdy} duplicates the results presented in \cite{SekharChivukula:2019qih} and \cite{Chivukula:2022tla}.} in \cite{deGiorgi:2020qlg} for brane localized scalars, with subsequent 
consequences for dark matter observables in \cite{deGiorgi:2021xvm,deGiorgi:2022yha}.\footnote{An erroneous calculation with a massive spin-2 KK particle as a freeze-in candidate was performed in \cite{Cai:2021nmk}, which was subsequently refuted in \cite{Gill:2023kyz} as a result of the Ward-identities of the theory.} \looseness=-1

Our computations elucidate the differences between the behavior of scattering amplitudes of matter in the bulk and localized on the brane, as well as the differences arising from the nature of matter (scalars, fermions or vector) and their various helicities. We will demonstrate that, for brane-localized matter, the anomalous growth in the scattering amplitudes only cancel in the case where the matter is localized to positions at the endpoints (the ``branes") of RS1. The cancellations we uncover are the result of the properties of the mode equations for the gravitational KK modes \cite{SekharChivukula:2019qih,SekharChivukula:2019yul,Chivukula:2020hvi,Chivukula:2022kju}, including the consequences of the $N=2$ SUSY structure relating the properties of the modes associated with the different helicities of the gravitational sector \cite{Lim:2007fy,Lim:2008hi}, as well as the mode equations for the matter particles.
In all cases we demonstrate that the residual amplitudes (after cancellations) grow no faster than $s/\Lambda_{\pi}^{2}$.

We also connect the observed cancellations to the underlying diffeomorphism invariance of the 5D gravitational theory. In what follows we will focus specifically on the scattering amplitudes for matter (modes of any helicity, arising either from either brane or bulk states) to produce longitudinally polarized spin-2 KK bosons.  It is these amplitudes which, due to polarization tensors of the external graviton KK modes, suffer from the largest potential energy growth. We show that the amplitudes for the production of longitudinal spin-2 KK states, after cancellation of the anomalous high-energy contributions from individual diagrams, can be interpreted using a ``KK Equivalence Theorem" analogous to the one in the compactified 5d KK Yang-Mills gauge theories~\cite{Chivukula:2001esy, Csaki:2003dt}.
In extra-dimensional gauge-theories the scattering amplitudes of the longitudinally polarized KK gauge bosons equals that of the corresponding KK Goldstone bosons in
the high energy limit. The power-counting of the scattering of Goldstone bosons, unlike those for massive KK gauge-states, is manifest, and has no anomalous high-energy growth. Specifically, in this paper we show that the leading non-vanishing contributions to the amplitudes in matter-gravity scattering involving longitudinal spin-2 states can be rewritten in terms of of the wavefunctions of the scalar gravitational KK Goldstone bosons (for arbitrary curvature) instead of those of the KK gravitons.\footnote{The form of the amplitudes can also be constructed via the double copy prescription, which we will also discuss in upcoming work.}\looseness=-1

 For gravity compactified on a torus, it has previously been shown that an equivalence theorem   can be established~\cite{Hang:2021fmp,Li:2022rel}, in which case the scattering amplitude of the longitudinally polarized KK gravitons equals that of corresponding gravitational scalar KK Goldstone bosons.  The results presented here suggest that the equivalence theorem can be extended to a warped geometry for the gravitation mode self-interactions and their interactions with matter.  A complete demonstration of the equivalence theorem in the RS1 model is beyond the scope of this paper, and is the subject of subsequent work~\cite{Chivukula:2023qrt}.   
 
 All of the potential bad high-energy behavior of the individual contributions to the scattering of longitudinal spin-2 KK states are, from the perspective of an equivalence theorem, just the usual naive unphysical high-energy behavior to be expected in a ``unitary gauge" calculation due to the unitary-gauge massive spin-2 propagators and external polarization states. This unphysical high-energy behavior of individual diagrams disappears in an `t-Hooft-Feynman-like gauge in which there are unphysical scalar (and, for gravity, vector) Goldstone states~\cite{Chivukula:2023qrt}. The connection between the cancellation of the high-energy growth of the scattering amplitudes demonstrated here and the diffeomorphism invariance of the underlying 5D gravitational theory is the ability to perform the analysis in either a unitary or an `t-Hooft-Feynman-like gauge, a freedom which relies on the diffeomorphism invariance of the underlying 5D gravitational theory. \looseness=-1

Finally, we will show that the sum-rules that ensure the cancellations of the anomalously growing contributions to the scattering amplitudes can be extended to models where the extra dimension is stabilized via the Goldberger-Wise mechanism \cite{Goldberger:1999uk}. Like the analogous calculation for spin-2 KK graviton self-interactions \cite{Chivukula:2021xod,Chivukula:2022tla,Chivukula:2022kju}, we will argue that the matter interactions within GW-stabilized model will involve additional contributions to the sum-rules from the GW scalars. \looseness=-1

The rest of the paper is organized as follows. In Section \ref{sec:Lag} we set up the gravitational Lagrangian, the metric, and the graviton sector mode expansions. In Section \ref{sec:matter}, we discuss matter-KK mode interactions for both bulk and brane matter. In Section \ref{sec:amp}, we describe the structure of the scattering amplitudes and the necessary sum-rules that ensure that scattering amplitudes are well behaved. We conclude in Section \ref{sec:conc}. We provide details of the calculation in the appendices for the interested reader. Appendix \ref{app:lag} gives the Lagrangian upto 4 point interactions between the gravity sector and matter for bulk and brane, relevant for scattering amplitude calculation. In appendix \ref{app:wfnbulk} we provide wave functions of gravitons and bulk matter while appendix \ref{app:coup} gives the coupling structures between brane/bulk matter and the gravity sector. Appendix \ref{sec:kinematics} gives our kinematic conventions and, finally appendix \ref{sec:sumrules} gives detailed proofs of sum-rules used in the main body of the paper. 
 \looseness=-1

\section{Gravitational Lagrangian, Metric, and Modes}
\label{sec:Lag}

The metric for the RS model in conformal coordinates $(x_{\mu},z)$ can be written as, 
\begin{equation}
    G_{MN} = e^{2A(z)}\begin{pmatrix}
        e^{-\kappa\hat{\varphi}/\sqrt{6}}(\eta_{\mu\nu}+\kappa\hat{h}_{\mu\nu}) & \frac{\kappa}{\sqrt{2}}\hat{A}_\mu \\
        \frac{\kappa}{\sqrt{2}}\hat{A}_\mu & -\left(1+\frac{\kappa}{\sqrt{6}}\hat{\varphi}\right)^2
    \end{pmatrix}~,
    \label{eq:background-metric}
\end{equation}
where the background 4D Minkowski metric $\eta_{\mu\nu}\equiv\rm Diag(+1,-1,-1,-1)$ is used to raise and lower indices. The line element is then written as, 
\begin{equation}
    ds^2 = e^{2A(z)}(\eta_{\mu\nu}dx^\mu dx^\nu - dz^2),
\end{equation}

The metric fluctuations $\hat{h}_{\mu\nu}(x,z)$ define the spin-2 fluctuations in 4D, while  the $\hat{A}_\mu$
and $\hat{\varphi}$ fields yield the spin-1 and spin-0 fluctuations respectively.
The warp factor $A(z)$, 
\begin{equation}
   A(z) = -\ln(kz)~,
\end{equation}
satisfies the Einstein equations for the bulk geometry,
\begin{equation}
    A''-(A')^2=0~,
    \label{eq:RS1EinsteinEq}
\end{equation}
and the value of the coupling $\kappa$ is set by the bulk and brane cosmological constants, such that the 4-dimensional Planck constant $M_{\rm Pl}$ is $\kappa_{4D}=2/M_{\rm Pl}$.
The extra-dimension spans the interval $z_1 \le z \le z_2$, where $z_1$ is the location of the ``Planck brane" and $z_2$  location of the ``TeV brane" respectively. 
The 5D RS Lagrangian can then be written as, 

\begin{equation}
    \mathcal{L}_{\rm 5D}^{\rm (RS)} = \mathcal{L}_{\rm EH} + \mathcal{L}_{\rm CC} + \Delta\mathcal{L}~,
\end{equation}
where $\mathcal{L}_{\rm EH}$ and $\mathcal{L}_{\rm CC}$ are the usual Einstein-Hilbert and cosmological constant terms respectively. The $\Delta\mathcal{L}$ term is a total derivative term required for a well defined variational principle for the action. 

The effective 4D action is obtained after KK decomposing the 5D field as \cite{Chivukula:2022kju}, 
\begin{eqnarray}
    \hat{h}_{\mu\nu}(x^\alpha,z) =&& \sum\limits_{n=0}^{\infty}\hat{h}_{\mu\nu}^{(n)}(x^\alpha)f^{(n)}(z),\label{eq:KK_1u}\\
    \hat{A}_{\mu}(x^\alpha,z) =&& \sum\limits_{n=1}^{\infty}\hat{A}_{\mu}^{(n)}(x^\alpha)g^{(n)}(z),\label{eq:KK_2u}\\
    \hat{\varphi}(x^\alpha,z) =&&~ \hat{r}(x^\alpha)k^{(0)}(z) +  \sum\limits_{n=1}^{\infty}\hat{\pi}^{(n)}(x)k^{(n)}(z)~,\label{eq:KK_3u}
\end{eqnarray}
and integrating over $z$. The massless graviton fields are given by $\hat{h}^{(0)}_{\mu\nu}$, while the massive KK graviton fields are $\hat{h}^{(n>0)}_{\mu\nu}$. The massless radion field is given by $\hat{r}$. The unphysical degrees of freedom, which can be eliminated using diffeomorphism invariance, are described by the spin-1 vector Goldstone modes $\hat{A}^{(n)}_\mu$ and the spin-0 scalar Goldstone modes $\hat{\pi}^{(n)}$. The wavefunctions satisfy the boundary conditions
\begin{equation}
    \partial_z f^{(n)}(z) = g^{(n)}(z) = \left[\partial_z+2A'(z)\right]k^{(n)}(z) = 0,~{\rm for}~z=z_{1,2}.
    \label{eq:bc_fgk}
\end{equation}

The details of the procedure to bring the Lagrangian to a canonical form, and the coupling structures of the 3- and 4-point vertices for the gravity sector been documented in \cite{Chivukula:2020hvi}.  In conformal coordinates,  the solutions to the Sturm–Liouville (SL) problems defining the modes subject to the boundary conditions are \cite{Chivukula:2022kju}, 
\begin{eqnarray}
    f^{(n)}(z) &=& C^{(n)}_h z^2\left[Y_1(m_n z_2)J_2(m_n z)-J_1(m_n z_2)Y_2(m_n z)\right],   \label{eq:spin2wfn} \\
    g^{(n)}(z) &=& C^{(n)}_A z^2\left[Y_1(m_n z_2)J_1(m_n z)-J_1(m_n z_2)Y_1(m_n z)\right], \\
    k^{(n)}(z) &=& C^{(n)}_\varphi z^2\left[Y_1(m_n z_2)J_0(m_n z)-J_1(m_n z_2)Y_0(m_n z)\right],
\end{eqnarray}
for the massive modes $n>0$, and
\begin{eqnarray}
    f^{(0)}(z) &=& C^{(0)}_h, \\
    g^{(0)}(z) &=& 0, \\
    k^{(0)}(z) &=& C^{(0)}_\varphi z^2,
    \label{eq:0modewfn}
\end{eqnarray}
for the massless modes, where $J_{a}$ and $Y_{a}$ are Bessel functions of the first and second kind, respectively. The normalizations $C^{(n)}_{h,A,\varphi}$ are fixed by
\begin{equation}
\aligned
    \int_{z_1}^{z_2}dz~e^{3A(z)}f^{(m)}(z)f^{(n)}(z) &= \int_{z_1}^{z_2}dz~e^{3A(z)}g^{(m)}(z)g^{(n)}(z)\\
    &= \int_{z_1}^{z_2}dz~e^{3A(z)}k^{(m)}(z)k^{(n)}(z) = \delta_{m,n}.
\endaligned
\label{eq:norm}
\end{equation}
where the spin-2 massless mode represents the usual massless 4D graviton that yields gravity in 4D, while the $k^{(0)}$ massless mode is the radion.
The mass $m_n$ of the KK gravitons is the $n$-th solution of the equation\footnote{The masses of the ``unphysical" vector and scalar states are degenerate with those for the physical spin-2 states as the result of an $N=2$ SUSY symmetry of the corresponding mode-equations \cite{Chivukula:2022kju}.}
\begin{equation}
    Y_1(m_n z_2)J_1(m_n z_1)-J_1(m_n z_2)Y_1(m_n z_1) = 0.
\end{equation}
The wavefunctions have an $N=2$ supersymmetric structure \cite{Lim:2007fy, Lim:2008hi,Chivukula:2022kju}, 
\begin{equation}
    \begin{cases}
        \partial_z f^{(n)} = m_ng^{(n)}\\
        (-\partial_z-3A') g^{(n)} = m_nf^{(n)}
    \end{cases}\qquad
    \begin{cases}
        (\partial_z+A') g^{(n)} = m_nk^{(n)}\\
        (-\partial_z-2A') k^{(n)} = m_ng^{(n)}~.
    \end{cases}
    \label{eq:susy_fgk}
\end{equation}
which we will use in what follows.

As mentioned previously, the Goldstone modes $\hat{A}^{(n)}_\mu$ and $\hat{\pi}^{(n)}$ can be gauged away \cite{Callin:2004zm}, and the relevant physical states are the spin-2 KK modes with wave function $f^{(n)}(z)$ starting from $n=0$ and a massless physical radial mode with wavefunction $k^{(0)}(z)$. In the rest of the paper, we work in such unitary gauge -- however, we will show that the leading non-zero scattering amplitudes involving helicity-0 spin-2 KK modes may be rewritten in terms of the the ``pion" wavefunctions $k^{(n)}(z)$ as expected from an equivalence theorem.

\section{Bulk and Brane matter}
\label{sec:matter}
In this section we lay out the relevant matter Lagrangians and interaction terms for matter coupled to gravity either in the brane or the bulk. Note that in contrast to previous papers \cite{SekharChivukula:2019qih,SekharChivukula:2019yul,Chivukula:2020hvi}, we work in conformal coordinates, and therefore the interaction Lagrangians, the Sturm-Liouville problem and the subsequent wavefunctions are defined in terms of these coordinates.

In the effective 4D description, 
the couplings of the spin-2 KK gravitons to matter (scalars, fermions or vectors) can be expressed by the following action,
\begin{equation}
    \mathcal{S}_{M}= \int d^{4}x~ \mathcal{L}(\tilde{G},s,v,f),
\end{equation}
which upon expanding to order $\kappa$ in the metric fluctuation yields, 
\begin{equation}
    \mathcal{S}_{M}= -\frac{\kappa}{2}\int d^{4}x~h_{\mu\nu}T^{\mu\nu}(s,v,f).
\end{equation}
The stress energy tensor $T_{\mu\nu}$ is given by, 
\begin{equation}
    T_{\mu\nu} = \left( -\eta_{\mu\nu}\mathcal{L} + 2\frac{ \delta \mathcal{L}}{\delta \tilde{G}^{\mu\nu}} \right)|_{\tilde{G}=\eta}.
\end{equation}

From this point onward the task is to compute scattering amplitudes of matter-KK mode interactions. We first lay out the relevant matter Lagrangians, and the corresponding 3- and 4-point interaction terms that will be used in the calculation of the scattering amplitudes.

\subsection{Brane Matter}
\begin{figure}
    \includegraphics[]{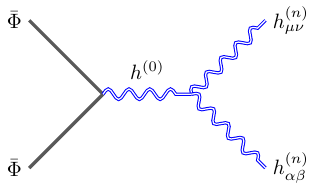}
    \includegraphics[]{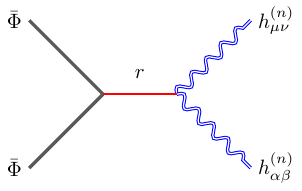}
   \includegraphics[]{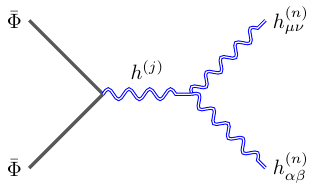}
   \includegraphics[]{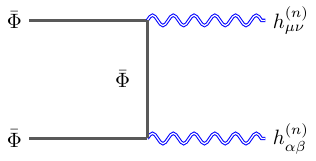}
    \includegraphics[]{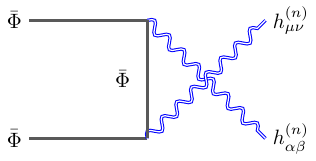}
    \includegraphics[]{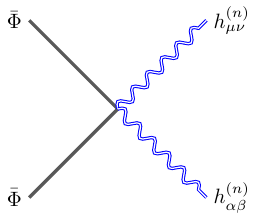}
     \caption{Brane localized matter (where $\bar{\Phi} = \bar{S}, \chi, \bar{V}$) annihilating to spin-2 KK modes. Here $r$ represents the radion.} 
    \label{fig:branescalar}
\end{figure}
We write the most general brane matter Lagrangian interacting with 
the spin-2 KK sector as, 

\begin{equation}
 \mathcal{L}_{\rm brane} = \mathcal{L}_{\rm spin-2} + \mathcal{L}_{M, \rm brane},
\end{equation}
where 
\begin{equation}
 \mathcal{L}_{M,\rm brane} = \mathcal{L}_{\bar{S},\rm brane} + \mathcal{L}_{\chi, \rm \rm brane} +\mathcal{L}_{\bar{V}, \rm brane},
 \end{equation}
 and $\mathcal{L}_{\bar{S}, \rm brane}$, $\mathcal{L}_{\chi, \rm brane} $ and  $\mathcal{L}_{\bar{V}, \rm brane}$ are the Lagrangian densities for brane localized scalars, fermions and vector fields respectively. 
The corresponding Lagrangians, localized on a brane at the boundaries $\bar{z} = z_1$~or~$z_2$, are given by

\begin{eqnarray}
    \mathcal{L}_{\bar{S}, \rm brane} & = & \int_{z_1}^{z_2}dz~\sqrt{\bar{G}}\left(\dfrac{1}{2}\bar{G}^{MN}\partial_M \bar{S}\partial_N \bar{S} - \dfrac{1}{2}M_{\bar{S}}^2 \bar{S}^2\right)e^{-2A(z)}\delta(z-\bar{z}), \\
     \mathcal{L}_{\chi, \rm brane} & = & \int_{z_1}^{z_2}dz~\sqrt{\bar{G}}\left(\bar{\chi}ie^{\mu}{}_{\bar{a}}\gamma^{\bar{a}}D_\mu\chi - M_\chi\bar{\chi}\chi\right)e^{-3A(z)}\delta(z-\bar{z}), \\
     \mathcal{L}_{\bar{V}, \rm brane} & = & \int_{z_1}^{z_2}dz\sqrt{\bar{G}}\left[-\dfrac{1}{4}\bar{G}^{MR}\bar{G}^{NS}\bar{F}_{MN}\bar{F}_{RS} + \dfrac{1}{2}M_{\bar{V}}^2\bar{G}^{MN}\bar{V}_M\bar{V}_N\right]\delta(z-\bar{z}).
\end{eqnarray}
The metric and its determinant is evaluated as an object induced on the brane enforced by the delta function.
Thus the brane localized quadratic kinetic term term can then be written in a canonically normalized form as
\begin{eqnarray}
  \mathcal{L}_{\bar{S}\bar{S}} & = & \dfrac{1}{2}\partial^\mu \bar{S}\partial_\mu \bar{S} - \dfrac{1}{2}m_{\bar{S}}^2 \bar{S}^2,\label{eq:branekinS}\\ 
 \mathcal{L}_{\chi\chi} & = & \left(i\bar{\chi}\slashed{\partial}\chi - m_\chi\bar{\chi}\chi\right), \label{eq:branekinF}\\  
 \mathcal{L}_{\bar{V}\bar{V}} & = & \dfrac{1}{2}\bar{V}^\mu\left[\eta_{\mu\nu}\left(\partial_\rho\partial^\rho + m_{\bar{V}}^2\right)-\left(1-\dfrac{1}{\xi}\right)\partial_\mu\partial_\nu\right]\bar{V}^\mu,
 \label{eq:branekinV}
\end{eqnarray}

For fermions, the covariant derivative on the fermion field is defined as 
\begin{equation}
    D_\mu\chi = \partial_\mu \chi + \frac{1}{2}\Omega_{\mu}{}^{\bar{a}\bar{b}}\sigma_{\bar{a}\bar{b}}\chi,
\end{equation}
where $\sigma_{\bar{a}\bar{b}} = [\gamma_{\bar{a}},\gamma_{\bar{b}}]/4$, with $\gamma_{\bar{a},\bar{b}}$ being the gamma matrices defined over the tetrad $e^{\nu\bar{a}}$. 
The induced  spin connection $\Omega_{\mu}{}^{\bar{a}\bar{b}}$ is given by
\begin{equation}
    \Omega_{\mu}{}^{\bar{a}\bar{b}} = e^{\nu\bar{a}}e_{\nu;\mu}{}^{\bar{b}} = e^{\nu\bar{a}}\left(\partial_\mu e_{\nu}{}^{\bar{b}} - e_{\rho}{}^{\bar{b}}\Gamma^{\rho}{}_{\mu\nu}\right).
\end{equation}
For vectors, in Eq. \ref{eq:branekinV} we have included a Proca mass term. While such mass term would break the 4D gauge symmetry, we will show that it does not spoil the unitarity for the scattering of $\bar{V}\bar{V}\rightarrow h^{(n)}h^{(n)}$, i.e diffeomorphism invariance in the gravity sector ensures that these processes are well behaved. In the case of massless gauge boson $M_{\bar{V}} = 0$, one would need to fix the gauge by the gauge-fixing term,
\begin{equation}
    \mathcal{L}_{\bar{V}, \rm GF} = \int_{z_1}^{z_2}dz~\left[-\dfrac{1}{2\xi}\left(\partial_\mu \bar{V}^\mu\right)^2\right]\delta(z-\bar{z}).
\end{equation}
which leads to the canonical Lagrangian in 4D given by, Eq. \ref{eq:branekinV}. 
Here we use reparametrized mass terms of the scalar, fermion and vector fields which are
\begin{eqnarray}
     m_{\bar{S}} & = & e^{A(\bar{z})}M_{\bar{S}}, \\
      m_{\chi} & = & e^{A(\bar{z})}M_{\chi}, \\
      m_{\bar{V}} & = & e^{A(\bar{z})}M_{\bar{V}}.
\end{eqnarray}

Note that unlike bulk fields, there are no interactions which contain an explicit derivative in the fifth dimension. We will show that this leads to different behaviours in the leading terms of matrix elements of the scattering amplitude calculations. From here-on we can perform the usual KK decomposition for the gravity sector to obtain an effective 4D action, with spin-2 KK graviton wave functions  given by Eq. \ref{eq:spin2wfn}. The 3- and 4-point interactions of the KK sector and matter are written out in Sections \ref{sec:couplingbscalar} - \ref{sec:couplingbvector}.

\subsection{Bulk Matter}
For matter in the bulk, we write the Lagrangian as, 
\begin{equation}
 \mathcal{L}_{M,\rm bulk} = \mathcal{L}_{\bar{S},\rm bulk} + \mathcal{L}_{\chi, \rm bulk} +\mathcal{L}_{\bar{V},\rm bulk},   
 \end{equation}

\begin{figure}
    \includegraphics[]{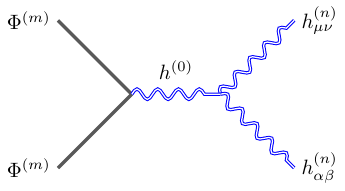}
    \includegraphics[]{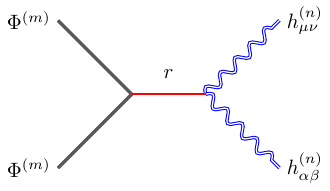}
   \includegraphics[]{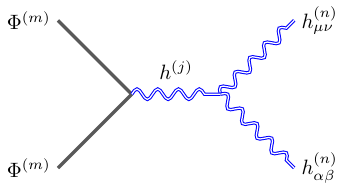}
   \includegraphics[]{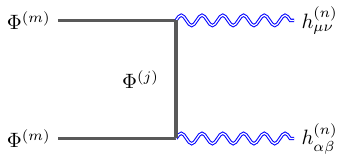}
    \includegraphics[]{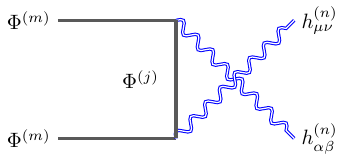}
    \includegraphics[]{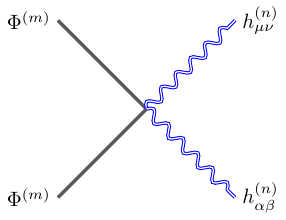}
     \caption{Bulk matter (where $\Phi = S, \psi, V$) annihilating to spin-2 KK modes. Note that, unlike brane matter shown in Fig. \protect\ref{fig:branescalar}, there are intermediate KK states which contribute in the $t$- and $u$-channels. Here $r$ represents the radion. }
    \label{fig:bulkscalar}
\end{figure}

The corresponding Lagrangians for a real bulk scalar $S$ with a mass $M_{S}$, a Dirac bulk (five-dimensional non-chiral) fermion 
\begin{equation}
    \psi=\begin{pmatrix} \psi_{L} \\ \psi_{R}  \end{pmatrix},
\end{equation}
with a bulk mass $M_{\psi}$, and a massless bulk gauge boson $V$ are given by, 
\begin{eqnarray}
    \mathcal{L}_{S, \rm bulk} & = & \sqrt{G}\left(\dfrac{1}{2}G^{MN}\partial_M S\partial_N S - \dfrac{1}{2}M_{S}^2 S^2\right), \\
    \mathcal{L}_{\psi,\rm bulk} &=&\sqrt{G}\left(\bar{\psi}iE^{M}{}_{a}\Gamma^aD_M\psi - M_\psi\bar{\psi}\psi\right), \\
     \mathcal{L}_{V,\rm bulk} & = & \sqrt{G}\left(-\dfrac{1}{4}F^{MN}F_{MN}\right).
\end{eqnarray}

Next we perform the integration over the extra-dimension 
and provide the canonical 4D Lagrangians for each of the species of matter. 
\begin{enumerate}
    \item {\bf{Scalars}: } Given the above scalar Lagrangian, the quadratic term  is canonically normalized as,
\begin{equation}
    \mathcal{L}_{SS} = \dfrac{1}{2}\int_{z_1}^{z_2}dz~e^{3A}\left\{\partial^\mu S\partial_\mu S - S\left[\left(-\partial_z-3A'\right)\partial_z + M_S^2e^{2A}\right]S\right\}.
\end{equation}
The bulk scalar can be decomposed into KK modes in the usual way,
\begin{equation}
    S(x^\alpha,z) = \sum_{n=0}^\infty S^{(n)}(x^\alpha)f^{(n)}_S(z),
\end{equation}
where $f^{(n)}_S$ are the eigenfunctions of the eigenequation
\begin{equation}
    \left[\left(-\partial_z-3A'\right)\partial_z + M_S^2e^{2A}\right]f^{(n)}_S(z) = m_{S,n}^2f^{(n)}_S(z).
\end{equation}
We choose the boundary condition to be\footnote{In principle, one could choose any Robin boundary conditions for the scalar wave functions $\partial_z S - \alpha_iS = 0 {~\rm at~z=z_i}$. Such a choice corresponds to adding  brane mass terms of the form $\Delta\mathcal{L}_S = \pm \alpha_{1,2} \sqrt{\bar{G}}~ e^{A}S^2\delta(z-z_{1,2})$. For simplicity, we choose the Neumann condition $\alpha_i = 0$.\looseness=-1},
\begin{equation}
    \partial_z f^{(n)}_S(z) = 0 \quad{\rm at }~z=z_{1,2}.
\end{equation}
Note that a massless mode exists only if $M_S =0$.
The corresponding wave functions and their orthogonality are provided in Appendix \ref{sec:wfnscalar}.

\item {\bf{Fermions}: } For fermions, we define the vierbein $E_{M}{}^{a}$ which satisfies,
\begin{equation}
    E_{M}^{a}E_{N}^{b}\eta_{ab} = G_{MN},
\end{equation}
and the gamma matrices in 5D defined by $ \Gamma^a = (\gamma^\mu, -i\gamma^5)$
such that they anti-commute, 
\begin{equation}
   \{\Gamma^a,\Gamma^b\}=2 \eta^{ab}.
\end{equation}
The covariant derivative on the fermion field is defined as,
\begin{equation}
    D_M\psi = \partial_M \psi + \frac{1}{2}\Omega_{M}{}^{ab}\sigma_{ab}\psi,
\end{equation}
where $\sigma_{ab} = [\Gamma_a,\Gamma_b]/4$, and the spin connection is given by
\begin{equation}
    \Omega_{M}{}^{ab} = E^{Na}E_{N;M}{}^{b} = E^{Na}\left(\partial_ME_{N}{}^{b} - E_{P}{}^{b}\Gamma^{P}{}_{MN}\right).
\end{equation}
In conformal coordinates, the quadratic term of the fermion Lagrangian can be written as,
\begin{equation}
    \mathcal{L}_\psi\supset e^{4A(z)}\left(\bar{\psi}_Li\slashed{\partial}\psi_L + \bar{\psi}_Ri\slashed{\partial}\psi_R 
 - \bar{\psi}_RD_\psi\psi_L - \bar{\psi}_LD_\psi^\dagger\psi_R \right),
\end{equation}
where the differential operator $D_\psi$ is defined as 
\begin{eqnarray}
    D_\psi &=& \partial_z +2A'(z)+M_\psi e^{A(z)},\\
    D_\psi^\dagger &=& -\partial_z -2A'(z)+M_\psi e^{A(z)}.
\end{eqnarray}
Note that $D_\psi^\dagger$ is the hermitian conjugate of $D_\psi$ with respect to the inner product
\begin{equation}
    \braket{g|f}_F = \int_{z_1}^{z_2}dz ~e^{4A(z)}g^*(z) f(z).
\end{equation}
After the compactification, the fermion fields can be expanded in KK modes as,
\begin{equation}
    \psi_{L/R}(x^\alpha,z) = \sum_n \psi^{(n)}_{L/R}(x^\alpha)f^{(n)}_{\psi_{L/R}}(z).
\end{equation}
where $f^{(n)}_{\psi_{L/R}}(z)$ are the wave functions of the left and right chiral fermions respectively. The wave functions satisfy the eigenequations
\begin{equation}
    \begin{cases}
        D_\psi f^{(n)}_{\psi_L} = m_{\psi, n}f^{(n)}_{\psi_R}~, \\
        D_\psi^\dagger f^{(n)}_{\psi_R} = m_{\psi, n}f^{(n)}_{\psi_L}~.
    \end{cases}
\end{equation}
with $m_{\psi, n}$ being the masses of the $n^{th}$ KK mode. Notice that the eigenequations are coupled, i.e they mix the left and the right handed sectors.
The mass spectra of $f^{(n)}_{\psi_L}$ and $f^{(n)}_{\psi_R}$ are degenerate, except for the zero mode, due to an $N=2$ quantum mechanical supersymmetry.  In order to have a massless left-handed fermion, one has to choose the boundary condition,
\begin{equation}
    D_\psi f^{(n)}_{\psi_L}(z)  = f^{(n)}_{\psi_R}(z) = 0 \quad{\rm at }~z=z_{1,2}.
\end{equation}
And the corresponding boundary condition for a massless right-handed fermion is,
\begin{equation}
    D_\psi f^{(n)}_{\psi_R}(z)  = f^{(n)}_{\psi_L}(z) = 0 \quad{\rm at }~z=z_{1,2}.
\end{equation}
The solutions to the eigenequations are the wave functions provided in Section. \ref{sec:wfnfermion} along with the corresponding orthonormality conditions.

\item {\bf{Vectors}: } For vectors, $F_{MN}$ is the 5D field strength tensor, 
\begin{equation}
    F_{MN} = \partial_MV_N-\partial_NV_M,
\end{equation} such that in conformal co-ordinates, the quadratic term of the gauge boson Lagrangian can be written as 
\begin{equation}
\aligned
    \mathcal{L}_{VV} = &~ \dfrac{1}{2}e^{A(z)}\left[V^\mu\left(\eta_{\mu\nu}\partial_\rho\partial^\rho - \partial_\mu\partial_\nu + \eta_{\mu\nu}(-\partial_z - A') \partial_z\right)V^\nu \right.\\
    &~ \left. - V_5\partial_\mu\partial^\mu V_5 + 2V_5\partial_\mu \partial_zV^\mu\right].
\endaligned
\end{equation}
The gauge fixing term is chosen to eliminate the terms involving mixing between $V_5$ and $ V^\mu$ in the above equation,
\begin{equation}
    \mathcal{L}_{V,\rm GF} = -e^{A}\dfrac{1}{2\xi}\left[\partial_\mu V^\mu - \xi e^{-A}\partial_z\left(e^A V_5\right)\right]^2.
\end{equation}
Then the gauge fixed quadratic terms become
\begin{equation}
\aligned
    \mathcal{L}_{VV+\rm GF} = &~ \dfrac{1}{2}e^{A(z)}\left[V^\mu\left(\eta_{\mu\nu}\partial_\rho\partial^\rho - \left(1-\dfrac{1}{\xi}\right)\partial_\mu\partial_\nu + \eta_{\mu\nu}D_V^\dagger D_V\right)V^\nu \right.\\
    &~ \left. \vphantom{\left(1-\dfrac{1}{\xi}\right)}- V_5\left(\partial_\mu\partial^\mu+\xi D_V D_V^\dagger\right) V_5 \right],
\endaligned
\end{equation}
where the differential operator $D_V$ is defined as 
\begin{eqnarray}
    D_V &=& \partial_z,\\
    D_V^\dagger &=& -\partial_z -A'(z).
\end{eqnarray}
Note that $D_V^\dagger$ is the hermitian conjugate of $D_V$ with respect to the inner product
\begin{equation}
    \braket{g|f}_V = \int_{z_1}^{z_2}dz ~e^{A(z)}g^*(z) f(z).
\end{equation}
After the KK compactification, the gauge boson fields can be expanded as
\begin{eqnarray}
    V_\mu(x^\alpha,z) &=& \sum_n V_\mu^{(n)}(x^\alpha)f^{(n)}_{V}(z),\\
    V_5(x^\alpha,z) &=& \sum_n V_5^{(n)}(x^\alpha)f^{(n)}_{V_5}(z).
\end{eqnarray}
The wavefunctions satisfy the eigenequations
\begin{equation}
    \begin{cases}
        D_V f^{(n)}_{V} = m_{V, n}f^{(n)}_{V_5}~ \\
        D_V^\dagger f^{(n)}_{V_5} = m_{V, n}f^{(n)}_{V}~
    \end{cases}.
\end{equation}
We choose the boundary condition to be,
\begin{equation}
    D_V f^{(n)}_{V} = f^{(n)}_{V_5} = 0 \quad{\rm at }~z=z_{1,2},
\end{equation}
such that $V_\mu$ has a massless mode and $V_5$ does not.
The solutions to the differential equations given in terms of eigenequations are given in Section. \ref{sec:wfnbulkvector}. 
\end{enumerate}

\section{Scattering amplitudes for Brane and Bulk Matter}
\label{sec:amp}

Consider the 2-to-2 elastic scattering of a pair of matter fields into a pair of longitudinally polarized KK gravitons, \begin{align}
    \bar{\Phi}_{\lambda}\bar{\Phi}_{\bar{\lambda}} \rightarrow h_{L}^{(n)}h_{L}^{(n)}, \quad \Phi^{(m)}_{\lambda}\Phi^{(m)}_{\bar{\lambda}}\rightarrow h_{L}^{(n)}h_{L}^{(n)},
\end{align}
where the $\bar{\Phi}$ represent incoming brane matter fields with $\bar{\Phi} = \bar{S},\chi,\bar{V}$, and $\Phi^{(m)}$ are bulk modes with $\Phi^{(m)} = S^{(m)},\psi^{(m)},V^{(m)}$; here $\lambda$, $\bar{\lambda}$ denote their helicities.
In the unitary gauge, there are 6 Feynman diagrams that contribute to the scattering amplitude,
\begin{equation}
    \mathcal{M}_{\lambda\bar{\lambda}} = \mathcal{M}_{tu,\lambda\bar{\lambda}} + \mathcal{M}_{h,\lambda\bar{\lambda}} + \mathcal{M}_{r,\lambda\bar{\lambda}} + \mathcal{M}_{4,\lambda\bar{\lambda}},
\end{equation}
where $\mathcal{M}_{tu,\lambda\bar{\lambda}}$ come from the $t$- and $u$-channel KK graviton or matter exchange diagrams, $\mathcal{M}_{h,\lambda\bar{\lambda}}$ corresponds to the $s$-channel diagrams with intermediate KK gravitons, $\mathcal{M}_{r,\lambda\bar{\lambda}}$ is the $s$-channel radion exchange contribution, and $\mathcal{M}_{4,\lambda\bar{\lambda}}$ comes from a 4-point contact interaction. These contributions are illustrated in Figs. \ref{fig:branescalar} and \ref{fig:bulkscalar} for brane matter and bulk matter.

To analyze the energy dependence of the scattering amplitude, we now expand the matrix element $\mathcal{M}_{\lambda\bar{\lambda}}$ in terms of the scattering energy $\sqrt{s}$ and the scattering angle $\theta$,
\begin{equation}
    \mathcal{M}_{\lambda\bar{\lambda}}(s,\theta) = \sum_{\sigma\in \mathbb{Z}}\widetilde{\mathcal{M}}^{(\sigma)}_{\lambda\bar{\lambda}}(\theta) s^{\sigma/2}.
\end{equation}
In the following sections we will analyze the energy growth of the scattering amplitudes for matter (brane or bulk) scattering into pairs of longitudinally polarized KK gravitons. We will determine the coefficients $\widetilde{\mathcal{M}}^{(\sigma)}_{\lambda\bar{\lambda}}(\theta)$ and demonstrate that the contributions for $\sigma >2$ vanish as the result of sum-rules which follow from the properties of the Sturm-Liouville problems for the mode expansions in the gravitation and matter sectors.

\subsection{Coupling structures}
In general, the self couplings of the entire compactified spin-2 sector, including
KK-gravitons and the radion couplings as well as any coupling of the KK sector with matter can be split up into two pieces due to the Lorentz structure, which we call $a$  and $b$ type couplings. The $a$ type couplings only have 4D derivatives, and therefore the overlap integrals contain only wave functions, while  $b$ couplings have derivatives over the compact dimension, such that overlap integrals contain explicit 5D derivatives. The structure of KK-sector self couplings was discussed in detail in \cite{Chivukula:2020hvi} and in conformal coordinates in \cite{Chivukula:2022kju}. Brane matter couplings to the KK sector involve only $a$ type couplings of the gravity sector since matter is confined to 4D brane. Here we describe the coupling structures that we will need for our calculations.

\subsubsection{Graviton self couplings}
The relevant self couplings within the gravitational sector are given by
\begin{eqnarray}
    a_{n_1n_2n_3} &=& \braket{f^{(n_1)}f^{(n_2)}f^{(n_3)}},\\
    b_{\bar{n}_1\bar{n}_2n_3} &=& \braket{(\partial_zf^{(n_1)})(\partial_zf^{(n_1)})f^{(n_3)}},\\
    b_{\bar{n}_1\bar{n}_2r} &=& \braket{(\partial_zf^{(n_1)})(\partial_zf^{(n_1)})k^{(0)}},
\end{eqnarray}
where the bracket $\braket{\cdots}$ denotes the inner product,
\begin{equation}
    \braket{ f_1^{(n_1)}f_2^{(n_2)}\cdots} = \int_{z_1}^{z_2}dz~e^{3A(z)} f_1^{(n_1)}(z)f_2^{(n_2)}(z)\cdots.
\end{equation}
\subsubsection{Couplings to matter fields}
The ``$a$-type" couplings between the matter fields and the graviton/radion fields, which contain no derivatives, are defined as,
\begin{eqnarray}
    a^{\Phi_1\Phi_2}_{n_1n_2n_3} &=& \braket{f^{(n_1)}f_{\Phi_1}^{(n_2)}f_{\Phi_2}^{(n_3)}}_{\Phi_1}, \\
    a^{\Phi_1\Phi_2}_{n_1n_2n_3n_4} &=& \braket{f^{(n_1)}f^{(n_2)}f_{\Phi_1}^{(n_3)}f_{\Phi_2}^{(n_4)}}_{\Phi_1},\\
    a^{\Phi_1\Phi_2}_{n_1n_2r} &=& \braket{f_{\Phi_1}^{(n_1)}f_{\Phi_2}^{(n_2)}k^{(0)}}_{\Phi_1},
\end{eqnarray}
where the bracket $\braket{\cdots}_\Phi$ denotes the inner product,
\begin{equation}
    \braket{ f_1^{(n_1)}f_2^{(n_2)}\cdots}_\Phi = \int_{z_1}^{z_2}dz~e^{w_\Phi A(z)} f_1^{(n_1)}(z)f_2^{(n_2)}(z)\cdots,\qquad 
    {\rm with~}\begin{cases}
        w_S = 3\\
        w_{\psi_L} = w_{\psi_R} = w_{\psi} = 4\\
        w_V = w_{V_5} = 1
    \end{cases}
\end{equation}
In the case of $\Phi_1 = \Phi_2$, we abbreviate the coupling as 
\begin{equation}
    a^{\Phi_1}_{\cdots} = a^{\Phi_1\Phi_1}_{\cdots}
\end{equation}
We also define the couplings that are related to the mass term in the Lagrangian as
\begin{eqnarray}
    a^{M_S}_{n\cdots n_1n_2} &=& \braket{e^{2A}f^{(n)}\cdots f_{S}^{(n_1)}f_{S}^{(n_2)}}_S, \\
    a^{M_\psi}_{n\cdots n_1n_2} &=& \braket{e^{A}f^{(n)}\cdots f_{\psi_L}^{(n_1)}f_{\psi_R}^{(n_2)}}_\psi, \\
    a^{M_S}_{n_1n_2r} &=& \braket{e^{2A} f_{S}^{(n_1)}f_{S}^{(n_2)}k^{(0)}}_S, \\
    a^{M_\psi}_{n_1n_2r} &=& \braket{e^{A} f_{\psi_L}^{(n_1)}f_{\psi_R}^{(n_2)}k^{(0)}}_\psi.
\end{eqnarray}
The ``$b$-type" couplings are defined in a similar manner as the ``$a$-type" couplings, except that we use a bar on top of the index to denote there is a derivative acting on the corresponding wavefunction,
\begin{equation}
    b^{\Phi_1\Phi_2}_{\cdots \bar{n}_i\cdots} = \braket{\cdots (\partial_zf_i^{(n_i)})\cdots}_\Phi
\end{equation}
A detailed account of the overlap integrals for 3- and 4-point interactions are provided in Sections \ref{sec:overlapscalar} - \ref{sec:overlapvector} along with the basic integration by parts and coupling identities.
\subsection{Amplitudes for brane-localized matter}
\subsubsection{Brane Scalar}
In the case of a brane localized scalar, the non-trivial contributions to the amplitude start at $\mathcal{O}(s^3)$, yielding a total
\begin{equation}
\aligned
    \widetilde{\mathcal{M}}^{(6)} =~& \frac{\kappa^2 (1-\cos2\theta)}{192 m_n^4}\left[\left(f^{(n)}(\bar{z})\right)^2 - \sum_{j=0}^{\infty}a_{nnj}f^{(j)}(\bar{z})\right],
\endaligned
\end{equation}
which vanishes due to completeness of the graviton wavefunctions,
\begin{equation}
\aligned
    \sum_{j=0}^{\infty}a_{nnj}f^{(j)}(\bar{z}) =~& \sum_{j=0}^{\infty}\left[\int_{z_1}^{z_2}dz~e^{3A(z)}f^{(n)}(z)f^{(n)}(z)f^{(j)}(z)\right]f^{(j)}(\bar{z}) \\
    =~& \int_{z_1}^{z_2}dz~f^{(n)}(z)f^{(n)}(z)\delta(z-\bar{z}) \\
    =~& \left[f^{(n)}(\bar{z})\right]^2.
\endaligned
\label{sr:brane_1}
\end{equation}
We note that the leading order ${\cal O}(s^3)$ amplitude vanishes independent of any condition on the brane $\bar{z}$. The situation changes at next order, as we demonstrate now.

At the order of $\mathcal{O}(s^2)$, after applying the sum rule above, the amplitude at next order can be written as,
\begin{equation}
\aligned
    \widetilde{\mathcal{M}}^{(4)} =~& -\frac{\kappa^2 }{576 m_n^4}\left\{(3\cos2\theta+1)\sum_{j=0}^{\infty}m_{j}^2a_{nnj}f^{(j)}(\bar{z}) +24b_{\bar{n}\bar{n}r}k^{(0)}(\bar{z})\right.\\
    &~\left.\vphantom{\sum_{j=0}^{\infty}}-2m_n^2(3\cos2\theta+5)\left[f^{(n)}(\bar{z})\right]^2  - 8m_n^2a_{nn0}f^{(0)}(\bar{z})~\right\}.
\endaligned
\end{equation}
One can use the eigenequations and the completeness relation to derive the following sum rule
\begin{equation}
\aligned
    \sum_{j=0}^{\infty}m_{j}^2a_{nnj}f^{(j)}(\bar{z}) =~& 2\sum_{j=0}^{\infty}(m_n^2a_{nnj}-b_{\bar{n}\bar{n}j})f^{(j)}(\bar{z})  =~ 2m_n^2\left[f^{(n)}(\bar{z})\right]^2,
\endaligned
\label{sr:brane_2}
\end{equation}
{\it but, as we explain below, the last equality relies on the fact that the wavefunctions $\partial_zf^{(n)} = g^{(n)}$ vanish at the location of the brane}, 
\begin{equation}
    \sum_{j=0}^{\infty}b_{\bar{n}\bar{n}j}f^{(j)}(\bar{z}) = m_n^2 \left[g^{(n)}(\bar{z})\right]^2 = 0.
\end{equation}
Using this relation the amplitude at sub-leading order becomes
\begin{equation}
\aligned
    \widetilde{\mathcal{M}}^{(4)} =~& -\frac{\kappa^2 }{72 m_n^4}\left\{3b_{\bar{n}\bar{n}r}k^{(0)}(\bar{z}) -m_n^2\left[f^{(n)}(\bar{z})\right]^2  - m_n^2a_{nn0}f^{(0)}(\bar{z})~\right\},
\endaligned
\end{equation}
which then vanishes due to the radion sum rule
\begin{equation}
\aligned
b_{\bar{n}\bar{n}r}k^{(0)}(\bar{z}) =~& \frac{m_n^2}{3}\left[f^{(n)}(\bar{z})\right]^2 + \frac{m_n^2}{3}a_{nn0}f^{(0)}(\bar{z}).
\endaligned
\label{sr:brane_3}
\end{equation}
The proof of the radion sum rule is given in Appendix~\ref{sec:sumrules}.

We emphasize that the cancellation of the bad ${\cal O}(s^2)$ high energy behavior crucially relies on the fact that the matter is localized at the boundaries $\bar{z} = z_1$~or~$z_2$, where the graviton KK mode wavefunctions satisfy $\partial_zf^{(n)}(\bar{z}) = g^{(n)}(\bar{z})=0$. The fact that the graviton wavefunctions have this property at the branes can be understood as the remnant of 5D diffeomorphism invariance. While the existence of the branes in RS breaks general 5D diffeomorphism invariance, the graviton Lagrangian is still invariant under the infinitesimal coordinate transformations that leave the location of the brane fixed
\begin{equation}
    x^M\mapsto \overline{x}^M=x^M+\xi^M.
    \label{eq:linear-diffeomorphisms}
\end{equation}
such that the parameter $\xi$ satisfies 
\begin{equation}
    \partial_z\xi_\mu(x^\alpha,z_i) = 0, \quad {\rm and} \quad \theta(x^\alpha,z_i) \equiv \xi^5(x^\alpha,z_i) = 0.
\end{equation}
As shown in \cite{Chivukula:2022kju}, the residual diffeomorphism is such that the parameters $\xi_\mu$ can be expanded in terms of the modes $f^{(j)}$, while the parameters $\theta$ have an $g^{(j)}$ mode-expansions.
Hence, for a ``translation" along the fifth dimension $\xi_\mu = 0$ and $\theta \neq 0$, the location of the brane matter at a fixed position is diffeomorphism invariant only if it is localized at the boundaries. Breaking such invariance would thus spoil the cancellation of the bad high energy behavior. For models with more than two branes, it is possible to localize the brane matter in the intermediate branes - but only if the appropriate boundary conditions are imposed in the gravitational sector - leading to a different form for the mode expansion and a different physical spectrum. The study of such scenario is beyond the scope of this work.

The residual non-vanishing amplitude starts at $\mathcal{O}(s)$. Applying all the previous sum rules, the leading non-zero contribution to the amplitude can be then written as,
\begin{equation}
    \widetilde{\mathcal{M}}^{(2)} =~ -\frac{\kappa^2 (3\cos2\theta+1)}{576 m_n^4}\left\{\sum_{j=0}^{\infty}m_{j}^4a_{nnj}f^{(j)}(\bar{z}) -2m_n^4\left[f^{(n)}(\bar{z})\right]^2 \right\}.
    \label{eq:residual_Sbar_1}
\end{equation}
Such expression can be further simplified, using the eigenequations, integration by part, and the fact that $A'' = (A')^2$ in the bulk, 
\begin{equation}
    \widetilde{\mathcal{M}}^{(2)} =~ -\frac{\kappa^2 (3\cos2\theta+1)}{96 }\left[f^{(n)}(\bar{z})\right]^2.
    \label{eq:residual_Sbar_3}
\end{equation}

Using the $N=2$ SUSY relations Eq.~(\ref{eq:susy_fgk}) and the boundary conditions Eq.~(\ref{eq:bc_fgk}), one can relate the KK graviton wavefunctions $f^{(n)}$ and scalar Goldstone wavefunctions $k^{(n)}$,
\begin{equation}
    k^{(j)}(\bar{z}) = -f^{(j)}(\bar{z}) -\dfrac{2A'(\bar{z})}{m_j}g^{(j)}(\bar{z}) = -f^{(j)}(\bar{z})\qquad({\rm for~}j> 0).
    \label{eq:f-to-k}
\end{equation}
Therefore, the amplitude can be written as 
\begin{equation}
    \widetilde{\mathcal{M}}^{(2)} =~ -\frac{\kappa^2 (3\cos2\theta+1)}{96 }\left[k^{(n)}(\bar{z})\right]^2~.\label{eq:residual_Sbar_2}
\end{equation}
We note that, while the amplitude in Eq.~(\ref{eq:residual_Sbar_1}) appears to be singular in the limit of $m_n\rightarrow0$, such singularity is not physical, as shown by Eqs.~(\ref{eq:residual_Sbar_3}) and (\ref{eq:residual_Sbar_2}). Another important observation is that Eq.~(\ref{eq:residual_Sbar_2}) depends solely on the wavefunction $k^{(n)}$ of the scalar Goldstone mode $\hat{\pi}^{(n)}$, consistent with what is expected from a Goldstone Equivalence Theorem~\cite{Chivukula:2023qrt}.

\subsubsection{Brane fermion}
For the scattering of brane fermions, the leading non-trivial contributions to the scattering amplitudes arise at $\mathcal{O}(s^3)$ and $\mathcal{O}(s^{5/2})$, depending on the helicity combinations chosen, and are given by 
\begin{eqnarray}
    \widetilde{\mathcal{M}}^{(6)}_{\pm\mp} &=~& \frac{\kappa^2 \sin2\theta}{192 m_n^4}\left[\left(f^{(n)}(\bar{z})\right)^2 - \sum_{j=0}^{\infty}a_{nnj}f^{(j)}(\bar{z})\right] = 0,\\
    \widetilde{\mathcal{M}}^{(5)}_{\pm\pm} &=~& \pm\frac{\kappa^2 m_\chi(1+\cos2\theta)}{96 m_n^4}\left[\left(f^{(n)}(\bar{z})\right)^2 - \sum_{j=0}^{\infty}a_{nnj}f^{(j)}(\bar{z})\right] = 0,
\end{eqnarray}
both of which vanish due the sum-rule  given in Eq.~(\ref{sr:brane_1}).

At next order the $\mathcal{O}(s^2)$ contributions also vanish
\begin{equation}
\aligned
    \widetilde{\mathcal{M}}^{(4)}_{\pm\mp} =~& \frac{\kappa^2 \sin2\theta}{192 m_n^4}\left\{\sum_{j=0}^{\infty}m_{j}^2a_{nnj}f^{(j)}(\bar{z}) -2m_n^2\left[f^{(n)}(\bar{z})\right]^2\right\} = 0,
\endaligned
\end{equation}
due to the sum-rule derived in Eq.~(\ref{sr:brane_2}). Again, it is crucial that the matter is localized at the boundaries.

The radion starts to contribute at $\mathcal{O}(s^{3/2})$, where its contribution to the amplitude at leading order can be written as
\begin{equation}
\aligned
    \widetilde{\mathcal{M}}^{(3)}_{\pm\pm} =~& \mp\frac{\kappa^2 }{72 m_n^4}\left\{3b_{\bar{n}\bar{n}r}k^{(0)}(\bar{z}) -m_n^2\left[f^{(n)}(\bar{z})\right]^2  - m_n^2a_{nn0}f^{(0)}(\bar{z})~\right\} = 0.
\endaligned
\end{equation}
and it vanishes due to the radion sum-rule given in Eq.~(\ref{sr:brane_3}).

The leading contribution to the residual amplitudes start at $\mathcal{O}(s)$ for helicities $\lambda\bar{\lambda} = \pm\mp$, and at $\mathcal{O}(s^{1/2})$ for helicities $\lambda\bar{\lambda} = \pm\pm$,
\begin{eqnarray}
    \widetilde{\mathcal{M}}^{(2)}_{\pm\mp} &=~& \frac{\kappa^2 \sin2\theta}{192 m_n^4}\left\{\sum_{j=0}^{\infty}m_{j}^4a_{nnj}f^{(j)}(\bar{z}) -2m_n^4\left[f^{(n)}(\bar{z})\right]^2 \right\},\\
    \widetilde{\mathcal{M}}^{(1)}_{\pm\pm} &=~& \pm\frac{\kappa^2 m_\chi(3\cos2\theta+1)}{288 m_n^4}\left\{\sum_{j=0}^{\infty}m_{j}^4a_{nnj}f^{(j)}(\bar{z}) -2m_n^4\left[f^{(n)}(\bar{z})\right]^2 \right\}.
\end{eqnarray}
Again, they can be simplified to a compact form of 
\begin{eqnarray}
    \widetilde{\mathcal{M}}^{(2)}_{\pm\mp} &=~& \frac{\kappa^2 \sin2\theta}{32 }\left[f^{(n)}(\bar{z})\right]^2=~ \frac{\kappa^2 \sin2\theta}{32 }\left[k^{(n)}(\bar{z})\right]^2,\\
    \widetilde{\mathcal{M}}^{(1)}_{\pm\pm} &=~& \pm\frac{\kappa^2 m_\chi(3\cos2\theta+1)}{48 }\left[f^{(n)}(\bar{z})\right]^2 =~ \pm\frac{\kappa^2 m_\chi(3\cos2\theta+1)}{48 }\left[k^{(n)}(\bar{z})\right]^2,
\end{eqnarray}
which are non-singular in the limit of $m_n\rightarrow 0$, leading to a form consistent with an equivalence theorem.

Note that for fermions, depending on whether a ``helicity flip" is required, the different spin channels have different power-counting behavior.

\subsubsection{Brane vector boson}
For the scattering of brane vector boson, the leading non-trivial contributions to the amplitudes for helicities $\lambda\bar{\lambda} = 00$ and $\pm\mp$ arise at $\mathcal{O}(s^3)$, for $\lambda\bar{\lambda} =\pm0/0\pm$ at $\mathcal{O}(s^{5/2})$, and for $\lambda\bar{\lambda} =\pm\pm$ at $\mathcal{O}(s^2)$, and are given by
\begin{eqnarray}
    \widetilde{\mathcal{M}}^{(6)}_{00} &=~& \widetilde{\mathcal{M}}^{(6)}_{\pm\mp} =~ \frac{\kappa^2 (\cos2\theta-1)}{192 m_n^4}\left[\left(f^{(n)}(\bar{z})\right)^2 - \sum_{j=0}^{\infty}a_{nnj}f^{(j)}(\bar{z})\right],\\
    \widetilde{\mathcal{M}}^{(5)}_{\pm0/0\pm} &=~& \pm\frac{\kappa^2 \sin2\theta}{48\sqrt{2} m_n^4}m_{\bar{V}}\left[\sum_{j=0}^{\infty}a_{nnj}f^{(j)}(\bar{z}) - \left(f^{(n)}(\bar{z})\right)^2\right],\\
    \widetilde{\mathcal{M}}^{(4)}_{\pm\pm} &=~& \frac{\kappa^2 (\cos2\theta+1)}{48 m_n^4}m_{\bar{V}}^2\left[ \sum_{j=0}^{\infty}a_{nnj}f^{(j)}(\bar{z}) - \left(f^{(n)}(\bar{z})\right)^2\right].
\end{eqnarray}
all of which vanish due the sum-rule  given in Eq.~(\ref{sr:brane_1}). We note that the amplitude $\widetilde{\mathcal{M}}^{(6)}_{\pm\pm}$ vanishes due to a direct cancellation between the $t$-, $u$-channel diagrams and the 4-point contact interaction, and it does not require any sum-rule.

The cancellation for helicities $\lambda\bar{\lambda} = \pm\mp$ at the subleading order $\mathcal{O}(s^2)$ further uses the sum-rule derived in Eq.~(\ref{sr:brane_2}),
\begin{equation}
\aligned
    \widetilde{\mathcal{M}}^{(4)}_{\pm\mp} =~& \frac{\kappa^2 (\cos2\theta-1)}{192 m_n^4}\left\{\sum_{j=0}^{\infty}m_{j}^2a_{nnj}f^{(j)}(\bar{z}) -2m_n^2\left[f^{(n)}(\bar{z})\right]^2\right\} = 0.
\endaligned
\end{equation}
The radion contributes to the scattering for the helicities $\lambda\bar{\lambda} = 00$ at $\mathcal{O}(s^2)$,
\begin{equation}
\aligned
    \widetilde{\mathcal{M}}^{(4)}_{00} =~& \frac{\kappa^2 }{72 m_n^4}\left\{3b_{\bar{n}\bar{n}r}k^{(0)}(\bar{z}) -m_n^2\left[f^{(n)}(\bar{z})\right]^2  - m_n^2a_{nn0}f^{(0)}(\bar{z})~\right\} = 0.
\endaligned
\end{equation}
which vanishes due to the radion sum-rule given in Eq.~(\ref{sr:brane_3}). 

At $\mathcal{O}(s^{3/2})$, the sub-amplitudes  
\begin{equation}
    \widetilde{\mathcal{M}}^{(3)}_{\pm0/0\pm} = 0
\end{equation}
vanish once the sum-rules in Eq.~(\ref{sr:brane_1}) and (\ref{sr:brane_2}) are applied.

Finally, similar to the behavior of brane scalars and fermions, the leading non-vanishing contribution to the amplitudes are at $\mathcal{O}(s)$ for $\lambda\bar{\lambda} = 00/\pm\mp$, and at $\mathcal{O}(s^{1/2})$ for $\lambda\bar{\lambda} = \pm0/0\pm$, and can be written as 
\begin{eqnarray}
    \widetilde{\mathcal{M}}^{(2)}_{00} &=~& \frac{\kappa^2 (3\cos2\theta+1)}{96 }\left[f^{(n)}(\bar{z})\right]^2=~ \frac{\kappa^2 (3\cos2\theta+1)}{96 }\left[k^{(n)}(\bar{z})\right]^2,\\
    \widetilde{\mathcal{M}}^{(2)}_{\pm\mp} &=~& \frac{\kappa^2 (\cos2\theta-1)}{32 }\left[f^{(n)}(\bar{z})\right]^2=~ \frac{\kappa^2 (\cos2\theta-1)}{96 }\left[k^{(n)}(\bar{z})\right]^2, \\
    \widetilde{\mathcal{M}}^{(1)}_{\pm 0/0\pm} &=~& \mp\frac{\kappa^2 \sin2\theta}{8\sqrt{2} }m_{\bar{V}}\left[f^{(n)}(\bar{z})\right]^2=~ \mp\frac{\kappa^2 \sin2\theta}{8\sqrt{2} }m_{\bar{V}}\left[k^{(n)}(\bar{z})\right]^2,
\end{eqnarray}
in a manner consistent with an equivalence theorem.

\subsection{Bulk scalar}

For the scattering of $m$-level KK scalar bosons to $n$-level KK gravitons, the non-trivial amplitude starts at $\mathcal{O}(s^3)$,
\begin{equation}
\aligned
    \widetilde{\mathcal{M}}^{(6)} =~& \frac{\kappa^2 }{192 m_n^4}\left[ (3\cos2\theta+5)\sum_{j=0}^{\infty}\left(a^S_{nmj}\right)^2 + (\cos2\theta-1)\sum_{j=0}^{\infty}a_{nnj}a^S_{jmm} - 4(\cos2\theta+1)a^S_{nnmm}\right],
\endaligned
\end{equation}
which vanishes due to completeness of the graviton  and scalar wavefunctions,
\begin{equation}
\sum_{j=0}^{\infty}\left(a^S_{nmj}\right)^2 = \sum_{j=0}^{\infty}a_{nnj}a^S_{jmm} = a^S_{nnmm}.
\label{sr:scalar_1}
\end{equation}

At the order of $\mathcal{O}(s^2)$, after applying the sum rule above, the amplitude at next order can be written as,
\begin{equation}
\aligned
    \widetilde{\mathcal{M}}^{(4)} =~& \frac{\kappa^2 }{192 m_n^4}\left\{(5-\cos2\theta)\sum_{j=0}^{\infty}m_{j}^2a_{nnj}a^S_{jmm} -2m_n^2(5-\cos2\theta)a^S_{nnmm}\right.\\
    &~\left.-2(\cos2\theta+3)\sum_{j=0}^{\infty}m_{S,j}^2\left(a^S_{nmj}\right)^2+2m_{S,m}^2(\cos2\theta+3)a^S_{nnmm}+16b^S_{\bar{n}\bar{n}mm}\right\}.
\endaligned
\end{equation}
One can use the eigenequations and the completeness relation to derive sum rules as
\begin{eqnarray}
    \sum_{j=0}^{\infty}m_{j}^2a_{nnj}a^S_{jmm}&=~&  2m_n^2a^S_{nnmm}-2b^S_{\bar{n}\bar{n}mm}, \label{sr:scalar_2}\\
    \sum_{j=0}^{\infty}m_{S,j}^2\left(a^S_{nmj}\right)^2&=~&  m_{S,m}^2a^S_{nnmm}+b^S_{\bar{n}\bar{n}mm}.\label{sr:scalar_3}
\end{eqnarray}
Once the above sum rules applied, the amplitude vanishes at this order,
\begin{equation}
    \widetilde{\mathcal{M}}^{(4)} = 0.
\end{equation}
It is interesting to note that, unlike other cases, the cancellation of the bad high energy for the bulk scalar case does not require the contribution from the radion, which only starts to appear at $\mathcal{O}(s)$.

The leading non-vanishing contribution to the amplitude starts at $\mathcal{O}(s)$. Applying all the previous sum rules, the residual amplitude can be then written as,
\begin{equation}
\aligned
   \widetilde{\mathcal{M}}^{(2)} =~& \frac{\kappa^2 }{576 m_n^4}\left\{24\sum_{j=0}^{\infty}m_{S,j}^4\left(a^S_{nmj}\right)^2 -(3\cos2\theta+1)\sum_{j=0}^{\infty}m_{j}^2a_{nnj}a^S_{jmm} \right.\\
    &~ + \left[2(3\cos2\theta+1)m_n^4+16m_n^2m_{S,m}^2-24m_{S,m}^4\right]a^S_{nnmm}\\
    &~ -8\left[(3\cos2\theta+1)m_n^2+4m_{S,m}^2\right]b^S_{\bar{n}\bar{n}mm} + 16m_n^2m_{S,m}^2a_{nn0}a^S_{0mm}\\
    &~ \left. \vphantom{\sum_{j=0}^{\infty}}- 144b_{\bar{n}\bar{n}r}\left(b^S_{\bar{m}\bar{m}r}+\dfrac{1}{3}M_S^2a^{M_S}_{mmr}\right)\right\}.
    \endaligned
\end{equation}
Although radion does not contribute to the cancellation, one can still derive the following radion sum rule, with details given in Appendix~\ref{sec:sumrules},
\begin{equation}
\aligned
    b_{\bar{n}\bar{n}r}\left(b^S_{\bar{m}\bar{m}r}+\dfrac{1}{3}M_S^2a^{M_S}_{mmr}\right) =&~ \dfrac{1}{9}m_n^2\left(m_{S,m}^2+3m_n^2\right)a^S_{nnmm} + \dfrac{1}{9}\left(7m_{S,m}^2-3m_n^2\right)b^S_{\bar{n}\bar{n}mm} \\
    &~ - \dfrac{2}{3}M_S^2b^{M_S}_{\bar{n}\bar{n}mm} + \dfrac{1}{9}m_n^2m_{S,m}^2a_{nn0}a^S_{0mm}\\
    &~ + \dfrac{10}{3}m_n^3\braket{A'f^{(n)}g^{(n)}f_S^{(m)}f_S^{(m)}}_S  + \dfrac{10}{3}m_n^2\braket{\left(A'\right)^2g^{(n)}g^{(n)}f_S^{(m)}f_S^{(m)}}_S.
\endaligned
\end{equation}
Together with another two sum-rules,
\begin{eqnarray}
    \sum_{j=0}^{\infty}m_{j}^4a_{nnj}a^S_{jmm}&=&~  8m_n^4a^S_{nnmm}-8m_n^2b^S_{\bar{n}\bar{n}mm}+24m_n^3\braket{A'f^{(n)}g^{(n)}f_S^{(m)}f_S^{(m)}}_S \\
    &&~ + 24m_n^2\braket{\left(A'\right)^2g^{(n)}g^{(n)}f_S^{(m)}f_S^{(m)}}_S,\\
    \sum_{j=0}^{\infty}m_{S,j}^4\left(a^S_{nmj}\right)^2&=&~  \left(m_{S,m}^4+3m_n^4\right)a^S_{nnmm}+2\left(3m_{S,m}^2-m_n^2\right)b^S_{\bar{n}\bar{n}mm}\\
    &&~ - 4M_S^2b^{M_S}_{\bar{n}\bar{n}mm}+24m_n^3\braket{A'f^{(n)}g^{(n)}f_S^{(m)}f_S^{(m)}}_S \\
    &&~ + 24m_n^2\braket{\left(A'\right)^2g^{(n)}g^{(n)}f_S^{(m)}f_S^{(m)}}_S,
\end{eqnarray}
and the fact that $k^{(n)} = -f^{(n)} - 2A'g^{(n)}/m_n$ (see Eq. (\ref{eq:f-to-k})),  the sub-amplitude can be written in a extremely compact form,
\begin{equation}
\aligned
    \widetilde{\mathcal{M}}^{(2)} =~& \dfrac{\kappa^2 (1-\cos2\theta)}{32}\braket{k^{(n)}k^{(n)}f_S^{(m)}f_S^{(m)}}_S,
\endaligned
\end{equation}
which is non-singular in the limit of $m_n\rightarrow 0$, and depends only on the wavefunctions of the scalar Goldstone boson $\hat{\pi}^{(n)}$, as expected from an equivalence theorem.

\subsection{Bulk fermion}
For the scattering of $m$-level bulk fermions to $n$-level gravitons, the non-trivial contributions to the amplitudes start at $\mathcal{O}(s^3)$, 
\begin{eqnarray}
    \widetilde{\mathcal{M}}^{(6)}_{-+} &=~& \frac{\kappa^2 \sin2\theta}{192 m_n^4}\left[3a^{\psi_L}_{nnmm} - 2\sum_{j=0}^{\infty}\left(a^{\psi_L}_{nmj}\right)^2 - \sum_{j=0}^{\infty}a_{nnj}a^{\psi_L}_{jmm}\right] =~ 0,\\
    \widetilde{\mathcal{M}}^{(6)}_{+-} &=~& \frac{\kappa^2 \sin2\theta}{192 m_n^4}\left[3a^{\psi_R}_{nnmm} - 2\sum_{j=0}^{\infty}\left(a^{\psi_R}_{nmj}\right)^2 - \sum_{j=0}^{\infty}a_{nnj}a^{\psi_R}_{jmm}\right] =~ 0.
\end{eqnarray}
both of which vanish due the completeness of the graviton and fermion wavefunctions,
\begin{equation}
\sum_{j=0}^{\infty}\left(a^{\psi_{L/R}}_{nmj}\right)^2 = \sum_{j=0}^{\infty}a_{nnj}a^{\psi_{L/R}}_{jmm} = a^{\psi_{L/R}}_{nnmm}.
\label{sr:fermion_1}
\end{equation}

At the order of $\mathcal{O}(s^{5/2})$, the amplitudes read,
\begin{equation}
    \widetilde{\mathcal{M}}^{(5)}_{\pm\pm} = \pm\frac{\kappa^2 (\cos2\theta+3)}{192 m_n^4}\left[m_{\psi,m}\left(a^{\psi_L}_{nnmm} + a^{\psi_R}_{nnmm}\right) - 2\sum_{j=0}^{\infty}m_{\psi,j} a^{\psi_L}_{nmj}a^{\psi_R}_{nmj}\right].
\end{equation}
One can use the eigenequations and the completeness relation to derive sum rules as
\begin{equation}
    2\sum_{j=0}^{\infty}m_{\psi,j} a^{\psi_L}_{nmj}a^{\psi_R}_{nmj} =  m_{\psi,m}a^{\psi_L}_{nnmm}+m_{\psi,m}a^{\psi_R}_{nnmm},
    \label{sr:fermion_2}
\end{equation}
which leads to vanishing sub-amplitudes at $\mathcal{O}(s^{5/2})$.

The amplitudes at the order of $\mathcal{O}(s^{2})$ can be written as,
\begin{eqnarray}
    \widetilde{\mathcal{M}}^{(4)}_{-+} &=~& \frac{\kappa^2 \sin2\theta}{192 m_n^4}\left[ 2\sum_{j=0}^{\infty}m_{\psi,j}^2\left(a^{\psi_L}_{nmj}\right)^2 + \sum_{j=0}^{\infty} m_j^2 a_{nnj}a^{\psi_L}_{jmm} - 2\left(m_{n}^2 + m_{\psi,m}^2\right) a^{\psi_L}_{nnmm} \right],\\
    \widetilde{\mathcal{M}}^{(4)}_{+-} &=~& \frac{\kappa^2 \sin2\theta}{192 m_n^4}\left[ 2\sum_{j=0}^{\infty}m_{\psi,j}^2\left(a^{\psi_R}_{nmj}\right)^2 + \sum_{j=0}^{\infty} m_j^2 a_{nnj}a^{\psi_R}_{jmm}  - 2\left(m_{n}^2 + m_{\psi,m}^2\right) a^{\psi_R}_{nnmm} \right],
\end{eqnarray}
which vanish once the following sum-rules are applied,
\begin{eqnarray}
    \sum_{j=0}^{\infty}m_{\psi,j}^2\left(a^{\psi_{L/R}}_{nmj}\right)^2 &=~&b^{\psi_{L/R}}_{\bar{n}\bar{n}mm} + m_{\psi,m}^2a^{\psi_{L/R}}_{nnmm},\label{sr:fermion_3}\\
    \sum_{j=0}^{\infty} m_j^2 a_{nnj}a^{\psi_{L/R}}_{jmm} &=&~2m_n^2a^{\psi_{L/R}}_{nnmm}  - 2b^{\psi_{L/R}}_{\bar{n}\bar{n}mm}.\label{sr:fermion_4}
\end{eqnarray}

The radion starts to contribute at $\mathcal{O}(s^{3/2})$, where the amplitude can be written as
\begin{equation}
\aligned
    \widetilde{\mathcal{M}}^{(3)}_{--} =~& -\frac{\kappa^2}{144 m_n^4}\left[ 6\sum_{j=0}^{\infty}m_{\psi,j}^3 a^{\psi_L}_{nmj}a^{\psi_R}_{nmj} - 2 m_{\psi,m}\left(b^{\psi_L}_{\bar{n}\bar{n}mm} + b^{\psi_R}_{\bar{n}\bar{n}mm}\right)\right. \\
    & +~m_{\psi,m}\left(m_n^2-3m_{\psi,m}^2\right)\left(a^{\psi_L}_{nnmm} + a^{\psi_R}_{nnmm}\right)\\
    & +~  m_n^2m_{\psi,m}a_{nn0}\left(a^{\psi_L}_{0mm} + a^{\psi_R}_{0mm}\right)\\
    & -~ \left. \vphantom{\sum_{j=0}^{\infty}}9m_{\psi,m}b_{\bar{n}\bar{n}r} \left(a^{\psi_L}_{mmr} + a^{\psi_R}_{mmr}-\frac{4M_\psi}{3m_{\psi,m}}a^{M_\psi}_{mmr}\right)\right]
\endaligned
\end{equation}
and it vanishes due to the sum-rules,
\begin{eqnarray}
    \sum_{j=0}^{\infty}m_{\psi,j}^3 a^{\psi_L}_{nmj}a^{\psi_R}_{nmj} &=~& \frac{3}{2}m_{\psi,m}\left(b^{\psi_{L}}_{\bar{n}\bar{n}mm} + b^{\psi_{R}}_{\bar{n}\bar{n}mm}\right) + \frac{1}{2}m_{\psi,m}^3\left(a^{\psi_{L}}_{nnmm} + a^{\psi_{R}}_{nnmm}\right) - 2M_\psi b^{M_\psi}_{\bar{n}\bar{n}mm},\label{sr:fermion_5}\\
    b_{\bar{n}\bar{n}r}\left(a^{\psi_L}_{mmr} + a^{\psi_R}_{mmr}-\frac{4M_\psi}{3m_{\psi,m}}a^{M_\psi}_{mmr}\right) &=~& \frac{7}{9}\left(b^{\psi_L}_{\bar{n}\bar{n}mm} + b^{\psi_R}_{\bar{n}\bar{n}mm}\right) + \frac{1}{9}m_n^2a_{nn0}\left(a^{\psi_L}_{0mm}+a^{\psi_R}_{0mm}\right)\nonumber\\
        &&+ \frac{1}{9}m_n^2\left(a^{\psi_L}_{nnmm}+a^{\psi_R}_{nnmm}\right)-\frac{4M_\psi}{3m_{\psi,m}}b^{M_\psi}_{\bar{n}\bar{n}mm}.\label{sr:fermion_6}
\end{eqnarray}
While the first sum-rule can be proved using the eigenequations and the completeness relation, the proof of the radion sum-rule on the second line also requires the completeness of the wavefunctions $\{k^{n}\}$ of the scalar Goldstone bosons \cite{Chivukula:2022kju}.

The non-vanishing helicity-violating residual amplitudes start at $\mathcal{O}(s)$,
\begin{eqnarray}
    \widetilde{\mathcal{M}}^{(2)}_{-+} &=~& \frac{\kappa^2 \sin2\theta}{192 m_n^4}\left[\sum_{j=0}^{\infty} m_j^4 a_{nnj}a^{\psi_L}_{jmm} - 2m_n^4a^{\psi_L}_{nnmm} + 8m_n^4b^{\psi_L}_{\bar{n}\bar{n}mm}\right],\\
    \widetilde{\mathcal{M}}^{(2)}_{+-} &=~& \frac{\kappa^2 \sin2\theta}{192 m_n^4}\left[\sum_{j=0}^{\infty} m_j^4 a_{nnj}a^{\psi_R}_{jmm} - 2m_n^4a^{\psi_R}_{nnmm} + 8m_n^4b^{\psi_R}_{\bar{n}\bar{n}mm}\right].
\end{eqnarray}
Again, they can be simplified to a compact form of 
\begin{eqnarray}
    \widetilde{\mathcal{M}}^{(2)}_{-+} &=~& \frac{\kappa^2 \sin2\theta}{32 }\braket{k^{(n)}k^{(n)}f_{\psi_L}^{(m)}f_{\psi_L}^{(m)}}_\psi,\\
    \widetilde{\mathcal{M}}^{(2)}_{+-} &=~& \frac{\kappa^2 \sin2\theta}{32 }\braket{k^{(n)}k^{(n)}f_{\psi_R}^{(m)}f_{\psi_R}^{(m)}}_\psi,
\end{eqnarray}
as is consistent with an equivalence theorem.

Similarly, the residual helicity-conserving amplitudes begin at order  $\mathcal{O}(s^{1/2})$, and can be written as
\begin{equation}
\aligned
    \widetilde{\mathcal{M}}^{(1)}_{\pm\pm} =~&  \mp\frac{\kappa^2m_{\psi,m}}{32 }\braket{k^{(n)}k^{(n)}(f_{\psi_L}^{(m)}f_{\psi_L}^{(m)}+f_{\psi_R}^{(m)}f_{\psi_R}^{(m)})}_\psi(\cos2\theta-5) \\
    &~ \mp\frac{\kappa^2 M_\psi}{3}\braket{e^Ak^{(n)}k^{(n)}f_{\psi_L}^{(m)}f_{\psi_R}^{(m)}}_\psi.
\endaligned
\end{equation}

\subsection{Bulk gauge bosons}

For the scattering of $m$-level bulk vector bosons to $n$-level gravitons, the non-trivial contributions to the amplitudes start at $\mathcal{O}(s^3)$, 
\begin{eqnarray}
    \widetilde{\mathcal{M}}^{(6)}_{00} &=~& \frac{\kappa^2 }{192 m_n^4}\left[ 4(\cos2\theta+1)a^{V_5}_{nnmm} - (3\cos2\theta+5)\sum_{j=0}^{\infty}\left(a^{V_5}_{nmj}\right)^2 - (\cos2\theta-1)\sum_{j=0}^{\infty}a_{nnj}a^{V_5}_{jmm}\right],\\
    \widetilde{\mathcal{M}}^{(6)}_{\pm\mp} &=~& \frac{\kappa^2 (\cos2\theta-1)}{192 m_n^4}\left[2a^{V}_{nnmm} - \sum_{j=0}^{\infty}\left(a^{V}_{nmj}\right)^2 - \sum_{j=0}^{\infty}a_{nnj}a^{V}_{jmm}\right],\\
    \widetilde{\mathcal{M}}^{(6)}_{\pm\pm} &=~& \frac{\kappa^2 (\cos2\theta+3)}{96 m_n^4}\left[a^{V}_{nnmm} - \sum_{j=0}^{\infty}\left(a^{V}_{nmj}\right)^2 \right].
\end{eqnarray}
All of them vanish due the completeness of the graviton and fermion wavefunctions,
\begin{eqnarray}
&&\sum_{j=0}^{\infty}\left(a^{V}_{nmj}\right)^2 = \sum_{j=0}^{\infty}a_{nnj}a^{V}_{jmm} = a^{V}_{nnmm},\label{sr:vector_1}\\
&&\sum_{j=0}^{\infty}\left(a^{V_5}_{nmj}\right)^2 = \sum_{j=0}^{\infty}a_{nnj}a^{V_5}_{jmm} = a^{V_5}_{nnmm}.
\label{sr:vector_2}
\end{eqnarray}

At order $\mathcal{O}(s^{5/2})$,
using the previous sum rules the amplitudes become,
\begin{equation}
    \widetilde{\mathcal{M}}^{(5)}_{0\pm/\pm0} = \pm\frac{\kappa^2 \sin2\theta}{96\sqrt{2} m_n^4}\left[2\sum_{j=0}^{\infty}m_{V,j} a^{V}_{nmj}a^{V_5}_{nmj} - m_{V,m}\left(a^{V}_{nnmm} + a^{V_5}_{nnmm}\right)\right],
\end{equation}
One can use the eigenequations and the completeness relation to derive a sum rule as
\begin{equation}
    \sum_{j=0}^{\infty}m_{V,j} a^{V}_{nmj}a^{V_5}_{nmj}=~  \dfrac{1}{2}m_{V,m}\left(a^{V}_{nnmm} + a^{V_5}_{nnmm}\right),
    \label{sr:vector_3}
\end{equation}
which leads to vanishing amplitudes
\begin{equation}
\aligned
    \widetilde{\mathcal{M}}^{(5)}_{0\pm/\pm0} =~& 0.
\endaligned
\end{equation}

At order $\mathcal{O}(s^2)$, the radion starts to contribute. The sub-amplitudes are given, after applying all the previous sum-rules, by
\begin{eqnarray}
    \widetilde{\mathcal{M}}^{(4)}_{00} &=~& \frac{\kappa^2 }{576 m_n^4}\left\{ 6(\cos2\theta-5)\sum_{j=0}^{\infty}m_{V,j}^2\left(a^{V_5}_{nmj}\right)^2 + (3\cos2\theta+1)\sum_{j=0}^{\infty}m_{j}^2a_{nnj}a^{V_5}_{jmm}\right.\\
    &&~~ - 8m_n^2a_{nn0}a^V_{0mm} + \dfrac{48b_{\bar{n}\bar{n}r}b^V_{\bar{m}\bar{m}r}}{m_{V,m}^2}+ 16m_{V,m}^2a^V_{nnmm}\\
    &&~~ - \left[2m_n^2(3\cos2\theta+5)+2m_{V,m}^2(3\cos2\theta-7)\right]a^{V_5}_{nnmm}\\
    &&~~ \left. + 16\sum_{j=1}^{\infty}\dfrac{m_n^2m_{V,m}^2}{m_j^2}a_{nnj}\left(a^{V_5}_{jmm}-a^{V}_{jmm}\right)\right\},\\
    \widetilde{\mathcal{M}}^{(4)}_{\pm\mp} &=~& \frac{\kappa^2 (\cos2\theta-1)}{192 m_n^4}\left[2\sum_{j=0}^{\infty}m_{V,j}^2\left(a^{V}_{nmj}\right)^2 +\sum_{j=0}^{\infty}m_j^2a_{nnj}a^{V}_{jmm} - 2(m_n^2+m_{V,m}^2)a^{V}_{nnmm}\right],\\ 
    \widetilde{\mathcal{M}}^{(4)}_{\pm\pm} &=~& \frac{\kappa^2}{72 m_n^4}\left[3\sum_{j=0}^{\infty}m_{V,j}^2\left(a^{V}_{nmj}\right)^2 - m_{V,m}^2a^{V}_{nnmm} - 2m_{V,m}^2a^{V_5}_{nnmm} - 3b_{\bar{n}\bar{n}r}a^V_{mmr}\right.\\
    &&~~
    \left. + 2\sum_{j=1}^{\infty}\dfrac{m_n^2m_{V,m}^2}{m_j^2}a_{nnj}\left(a^{V_5}_{jmm}-a^{V}_{jmm}\right)\right].
\end{eqnarray}
One can use the eigenequations and the completeness relation to derive the following sum rules
\begin{eqnarray}
    \sum_{j=0}^{\infty}m_{V,j}^2\left(a^{V_5}_{nmj}\right)^2&=&~ b^{V_5}_{\bar{n}\bar{n}mm} + m_{V,m}^2a^{V_5}_{nnmm},\\
    \sum_{j=0}^{\infty}m_{j}^2a_{nnj}a^{V_5}_{jmm}&=&~  2m_n^2a^{V_5}_{nnmm} - 2 b^{V_5}_{\bar{n}\bar{n}mm}.
\end{eqnarray}
With the help of the completeness of $\{k^{n}\}$, one can derive the following radion sum rules, 
\begin{eqnarray}
    b_{\bar{n}\bar{n}r}b^V_{\bar{m}\bar{m}r} &=~& \dfrac{2}{3}m_{V,m}^2 b^{V_5}_{\bar{n}\bar{n}mm} - \dfrac{1}{3}m_{V,m}^4a^V_{nnmm}+\dfrac{1}{6}m_{V,m}^2(m_n^2+2m_{V,m}^2)a^{V_5}_{nnmm} \\
    &&~~ +\dfrac{1}{6}m_n^2m_{V,m}^2a_{nn0}a^V_{0mm} - \dfrac{1}{3} \sum_{j=1}^{\infty}\dfrac{m_n^2m_{V,m}^4}{m_j^2}a_{nnj}\left(a^{V_5}_{jmm}-a^{V}_{jmm}\right),\\
    b_{\bar{n}\bar{n}r}a^V_{mmr} &=~& b^{V}_{\bar{n}\bar{n}mm} + \dfrac{2}{3}m_{V,m}^2\left(a^V_{nnmm}-a^{V_5}_{nnmm}\right) + \dfrac{2}{3} \sum_{j=1}^{\infty}\dfrac{m_n^2m_{V,m}^2}{m_j^2}a_{nnj}\left(a^{V_5}_{jmm}-a^{V}_{jmm}\right)
\end{eqnarray}
And thus the total amplitudes also vanish at $\mathcal{O}(s^2)$,
\begin{equation}
    \widetilde{\mathcal{M}}^{(4)}_{00} = \widetilde{\mathcal{M}}^{(4)}_{\pm\mp} = \widetilde{\mathcal{M}}^{(4)}_{\pm\pm} = 0.
\end{equation}

At the order of $\mathcal{O}(s^{3/2})$, the radion does not contribute and no new sum-rules are need. The sub-amplitudes vanish once all the previous sum-rules are applied,
\begin{equation}
    \widetilde{\mathcal{M}}^{(3)}_{0\pm/\pm0} = 0.
\end{equation}

The non-vanishing amplitudes start at $\mathcal{O}(s)$, and may be written
\begin{eqnarray}
    \widetilde{\mathcal{M}}^{(2)}_{00} &=&~ \frac{\kappa^2 (3\cos2\theta+13))}{96}\braket{k^{(n)}k^{(n)}f^{(m)}_{V_5}f^{(m)}_{V_5}}_V,\\
    \widetilde{\mathcal{M}}^{(2)}_{\pm\mp} &=&~ \frac{\kappa^2 (3\cos2\theta+1))}{96}\braket{k^{(n)}k^{(n)}f^{(m)}_{V}f^{(m)}_{V}}_V,\\
    \widetilde{\mathcal{M}}^{(2)}_{\pm\pm} &=&~ 0.
\end{eqnarray}
At the order of $\mathcal{O}(s^{1/2})$, the leading amplitudes can be written as
\begin{eqnarray}
    \widetilde{\mathcal{M}}^{(1)}_{\pm 0/0\pm} &=&~ \pm\frac{\kappa^2 (3\cos2\theta-11))\cot\theta}{48\sqrt{2}}m_{V,m}\braket{k^{(n)}k^{(n)}(f^{(m)}_{V}f^{(m)}_{V}+f^{(m)}_{V_5}f^{(m)}_{V_5})}_V
\end{eqnarray}
Note again that all forms are consistent with the expectations from an equivalence theorem.

\section{Scattering amplitudes with a Goldberger-Wise stabilized geometry}
\label{sec:GWamp}

While all the results above are derived for an unstabilized RS1 model, they can be easily generalized to the case in which the size of the extra dimension is dynamically stabilized via the Goldberger-Wise mechanism.
The Goldberger-Wise mechanism \cite{Goldberger:1999uk} introduces a bulk scalar field $\hat{\Phi}$ with the kinetic term and potential terms
\begin{eqnarray}
    \mathcal{L}_{\Phi\Phi} &=& \sqrt{G}\left[\frac{1}{2} G^{MN} \partial_M\hat{\Phi}\partial_N\hat{\Phi}\right]~,\\
    \mathcal{L}_{\rm pot} &=& -\frac{4}{\kappa^2}\left[\sqrt{G}V[\hat{\Phi}] + \sqrt{\overline{G}}V_1[\hat{\Phi}]\delta_1(z-z_1)+ \sqrt{\overline{G}}V_2[\hat{\Phi}]\delta_1(z-z_2) \right]~.
\end{eqnarray}
The potential terms are chosen such that the ground state has a non-zero $z$-dependent expectation
value for $\hat{\Phi}$, and such that minimizing the action fixes the proper length of the extra dimension.
The bulk scalar field $\hat{\Phi}$ can be expanded around the background as 
\begin{equation}
    \hat{\Phi}(x^\alpha,z) = \frac{1}{\kappa} (\phi_0(z) + \hat{\phi}(x^\alpha,z)).
\end{equation}
Under the assumption that the GW scalar $\hat{\Phi}$ is a part of the gravity sector and does not directly couple to the matter fields,  the GW scalar only contributes to the scattering via its mixing with the radion.~\footnote{In general, $\hat{\Phi}$ could directly couple to the matter fields, and such interactions would contribute to the scattering amplitudes in a model-dependent fashion. The analysis in such cases is beyond the scope of this work.}

Following the notation in Ref.~\cite{Chivukula:2022kju}, the GW sector can be decomposed as,
\begin{equation}
    \hat{\Psi}(x^\alpha,z) = \sum\limits_{n=1}^{\infty}\hat{\pi}^{(n)}(x^\alpha)K^{(n)}(z) + \sum\limits_{n=0}^{\infty}\hat{r}^{(n)}(x)\tilde{K}^{(n)}(z),
\end{equation}
where 
\begin{equation}
    \hat{\Psi}(x^\alpha,z) = \begin{pmatrix}
        \hat{\varphi}(x^\alpha,z)\\\hat{\phi}(x^\alpha,z)\end{pmatrix},\quad
    K^{(n)}(z) = \begin{pmatrix}
        k^{(n)}(z)\\l^{(n)}(z)
    \end{pmatrix},\quad
    \tilde{K}^{(n)}(z) = \begin{pmatrix}
        \tilde{k}^{(n)}(z)\\\tilde{l}^{(n)}(z)
    \end{pmatrix}.
\end{equation}
The Goldstone modes $\hat{\pi}^{(n)}$ are rotated away in the unitary gauge, and the physical scalars $\hat{r}^{(n)}$ now replace the role of the radion to unitarize the scattering amplitudes. In particular, the completeness of the wavefunctions $\{k^{(n)}\}$ is modified,
\begin{equation}
     \xi(z') = \sum\limits_{n=1}^{\infty}k^{(n)}(z')\braket{k^{(n)}\xi}+\sum\limits_{n=0}^{\infty}\tilde{k}^{(n)}(z')\braket{\tilde{k}^{(n)}\xi}~,
\label{eq:k-special-completeness}
\end{equation}
in comparison to the one in the unstabilized RS1 model,
\begin{equation}
     \xi(z') = \sum\limits_{n=0}^{\infty}k^{(n)}(z')\braket{k^{(n)}\xi},
\label{eq:k-completeness}
\end{equation}
where $\xi(z')$ is an arbitrary function that satisfies the proper boundary conditions. 

To generalize the radion sum-rules discussed above to the GW model, we consider the Feynman diagrams of exchanging the physical GW scalars. At leading order in $s$, the masses of the GW scalars can be neglected. Thus, the scattering amplitude can be obtained by simply replacing the radion wavefunction $k^{(0)}$ in the RS1 by a tower of the GW scalar wavefunction $\tilde{k}^{(i)}$.
Therefore, we can generalize the radion sum-rules to the GW model by replacing all the radion-related couplings by the couplings involving the physical scalars $\hat{r}^{(i)}$,
\begin{equation}
    \braket{\cdots k^{(0)}}\braket{k^{(0)}\cdots } \Rightarrow \sum_{i=0}^{\infty}\braket{\cdots \tilde{k}^{(i)}}\braket{\tilde{k}^{(i)}\cdots }.
\end{equation}
We note that such generalization at leading order is sufficient for all the radion sum-rules given in this paper, because the radion contribution only appears at the lowest non-trivial order of the cancellation. For the residual terms at $\mathcal{O}(s)$ and below, they receive an additional contribution that is proportional to the masses of the scalar fields $\hat{r}^{(i)}$, which cannot be deduced from the scattering amplitudes involving a massless radion in RS1. An example is the scattering amplitudes of four KK gravitons. As shown in Ref.~\cite{Chivukula:2021xod, Chivukula:2022tla}, the leading order radion contribution appears at $\mathcal{O}(s^3)$, where the radion sum-rules can be generalized as above. But the cancellation of the scattering amplitude at order of $\mathcal{O}(s^2)$ requires an additional radion sum-rule that contains terms proportional to the scalar masses $\mu^2_{(i)}$, as in Eq.~(22) in Ref.~\cite{Chivukula:2022tla}. \looseness=-1

\section{Conclusion}
\label{sec:conc}

In this paper we have performed a comprehensive analysis of the scattering of matter and gravitational Kaluza-Klein modes in compactified five-dimensional gravity theories. We considered the scattering amplitudes for matter localized on a brane as well as in the bulk of the extra dimension for
scalars, fermions and vectors respectively, and considered an arbitrary warped RS background. 
While naive power-counting suggests that these amplitudes could grow as fast as ${\cal O}(s^3)$ [where $s$ is the center-of-mass scattering energy-squared], we demonstrated by explicit computation that cancellations between the various contributions result in a total amplitude which grows no faster than ${\cal O}(s)$. 

Extending previous work on the self-interactions of the gravitational KK modes, we showed that these cancellations occur due to sum-rule relations between the couplings and the masses of the modes that can be proven from the properties of the mode equations describing the gravity and matter wavefunctions. We demonstrated that these properties are tied to  the underlying diffeomorphism invariance of the five-dimensional theory.  We showed how our results generalize when the size of the extra-dimension is stabilized via the Goldberger-Wise mechanism. Our results show that naive calculations \cite{Garny:2015sjg,Lee:2013bua,Folgado:2019sgz} of the freeze-out and freeze-in relic abundance calculations for dark matter models including a spin-2 portal arising from an underlying five-dimensional theory will yield incorrect results.

Our computations further showed that the form of the leading high-energy behavior of graviton-matter KK scattering with external helicity-0 spin-2 states has the form expected from a gravitational equivalence theorem analogous to one in compactified 5d Yang-Mills gauge theory 
\cite{Chivukula:2001esy, Csaki:2003dt}, namely that the leading non-zero amplitudes are proportional to overlap integrals involving the wavefunctions of the scalar gravitational KK Goldstone bosons. In future work \cite{Chivukula:2023qrt} we will prove the gravitational equivalence theorem, which has been established for the self-interactions of the gravitational modes in toroidal compactification \cite{Hang:2021fmp,Li:2022rel}, generalizes to warped geometry and also to the interaction of gravity and matter modes as expected from the results reported here.

\bigskip

\noindent {\bf Acknowledgements:} The authors thank Dennis Foren for collaboration during the initial stages of this work. The work of RSC, EHS, and XW was supported in part by the National Science Foundation under Grant No. PHY-2210177. JAG acknowledges the support he has
received for his research through the provision of an Australian Government Research Training Program Scholarship.
Support for this work was provided by the University of Adelaide and the Australian Research Council through the Centre of Excellence for Dark Matter Particle Physics (CE200100008). The work of KM was supported in part by the National Science Foundation under Grant No. PHY-2310497. JAG and DS thank Anthony G. Williams for fruitful discussions. 


\appendix
\section{Lagrangian}
\label{app:lag}
In this appendix, we give the relevant Lagrangian up to 4-point interactions.
\subsection{Brane Matter}
\subsubsection{ Scalar}
\label{sec:couplingbscalar}
The 3-point interactions are given by the Lagrangian
\begin{eqnarray}
    \mathcal{L}_{h\bar{S}\bar{S}} &=& \dfrac{\kappa}{2}\int_{z_1}^{z_2}dz~ \hat{h}^{\mu\nu}\left[-\partial_\mu\bar{S}\partial_\nu\bar{S}+\dfrac{1}{2}\eta_{\mu\nu}\left(\partial_\rho\bar{S}\partial^\rho\bar{S}-m_{\bar{S}}^2\bar{S}^2\right)\right]\delta(z-\bar{z}),\\
    \mathcal{L}_{\varphi \bar{S}\bar{S}} &=& \dfrac{\kappa}{2}\int_{z_1}^{z_2}dz~ \frac{1}{\sqrt{6}}\hat{\varphi}\left[-\partial^\mu\bar{S}\partial_\mu\bar{S}+2m_{\bar{S}}^2\bar{S}^2\right]\delta(z-\bar{z}).
\end{eqnarray}
The 4-point interactions are given by the Lagrangian
\begin{eqnarray}
    \mathcal{L}_{hh\bar{S}\bar{S}} &=& \dfrac{\kappa^2}{4}\int_{z_1}^{z_2}dz~ \left[\vphantom{\frac{1}{4}}\left(2\hat{h}^{\mu\rho}\hat{h}^\nu{}_\rho-\hat{h}\hat{h}^{\mu\nu}\right)\partial_\mu\bar{S}\partial_\nu\bar{S}\right.\nonumber\\
    &&~+\left.\dfrac{1}{4}\left(\hat{h}^2-2\hat{h}^{\nu\rho}\hat{h}_{\nu\rho}\right)\left(\partial_\rho\bar{S}\partial^\rho\bar{S}-m_{\bar{S}}^2\bar{S}^2\right)\right]\delta(z-\bar{z}),\\
    \mathcal{L}_{\varphi\varphi \bar{S}\bar{S}} &=& \dfrac{\kappa^2}{4}\int_{z_1}^{z_2}dz~ \dfrac{1}{6}\hat{\varphi}^2\left(\partial^\mu\bar{S}\partial_\mu\bar{S}-4m_{\bar{S}}^2\bar{S}^2\right)\delta(z-\bar{z}),\\
    \mathcal{L}_{h\varphi \bar{S}\bar{S}} &=& \dfrac{\kappa^2}{2}\int_{z_1}^{z_2}dz~ \dfrac{1}{\sqrt{6}}\hat{\varphi}\hat{h}^{\mu\nu}\left[\partial_\mu\bar{S}\partial_\nu\bar{S}\vphantom{\dfrac{1}{2}}\right. \nonumber\\
    &&~\left.-\eta_{\mu\nu}\left(\dfrac{1}{2}\partial^\rho\bar{S}\partial_\rho\bar{S}-m_{\bar{S}}^2\bar{S}^2)\right)\right]\delta(z-\bar{z}).
\end{eqnarray}

\subsubsection{Brane fermion}
\label{sec:couplingbfermion}
The 3-point interactions are given by the Lagrangian
\begin{eqnarray}
    \mathcal{L}_{h\bar{\chi}\chi} &=& \kappa\int_{z_1}^{z_2}dz \left\{\dfrac{1}{4}\hat{h}^{\mu\nu}\left[\bar{\chi}\left(-i\gamma_\mu\overset{\leftrightarrow}{\partial}_\nu + i\eta_{\mu\nu}\overset{\leftrightarrow}{\slashed{\partial}}\right)\chi\right]\right.\\
    &&~~~\left.-\dfrac{1}{2}M_\chi e^A\hat{h}\bar{\chi}\chi\right\}\delta(z-\bar{z}),\\
    \mathcal{L}_{\varphi\bar{\chi}\chi} &=& \kappa\int_{z_1}^{z_2}dz \left\{\dfrac{3}{4\sqrt{6}}\hat{\varphi}\left[-i\bar{\chi}\overset{\leftrightarrow}{\slashed{\partial}}\chi \right] +~\dfrac{2}{\sqrt{6}}M_\chi e^A\hat{\varphi}\bar{\chi}\chi\right\}\delta(z-\bar{z}),
\end{eqnarray}
where the derivative $\overset{\leftrightarrow}{\partial}$ acts only on the fermion fields and is defined as
\begin{equation}
    \overset{\leftrightarrow}{\partial}_M = \overset{\rightarrow}{\partial}_M - \overset{\leftarrow}{\partial}_M.
\end{equation}
The 4-point interactions are given by the Lagrangian
\begin{eqnarray}
    \mathcal{L}_{hh\bar{\chi}\chi} &=& \dfrac{\kappa^2}{2}\int_{z_1}^{z_2}dz \left\{\dfrac{1}{8}\left(3\hat{h}^{\mu\rho}\hat{h}^\nu{}_\rho-2\hat{h}\hat{h}^{\mu\nu}\right)\left(i\bar{\chi}\gamma_\mu\overset{\leftrightarrow}{\partial}_\nu\chi\right)\right.\nonumber\\
    &&~+\dfrac{1}{8}\left(\hat{h}^2-2\hat{h}^{\nu\rho}\hat{h}_{\nu\rho}\right)\left[i\bar{\chi}\overset{\leftrightarrow}{\slashed{\partial}}\chi - 2M_\chi e^A\bar{\chi}\chi\right]\\
    &&~\left. + \dfrac{1}{8}\epsilon^{\lambda\alpha\beta\rho}h^\mu{}_\alpha\partial_\rho h_{\mu\beta}\left[\bar{\chi}_L\gamma_\lambda \chi_L-(L\leftrightarrow R)\right]\right\}\delta(z-\bar{z})\\
    \mathcal{L}_{\varphi\varphi \bar{\chi}\chi} &=& -\dfrac{\kappa^2}{2}\int_{z_1}^{z_2}dz ~\hat{\varphi}^2\left(\dfrac{3}{16}i\bar{\chi}\overset{\leftrightarrow}{\slashed{\partial}}\chi + \dfrac{2}{3}m_\chi\bar{\chi}\chi\right)\delta(z-\bar{z}),\\
    \mathcal{L}_{h\varphi \bar{\chi}\chi} &=& \kappa^2\int_{z_1}^{z_2}dz~ \dfrac{3}{8\sqrt{6}}\hat{\varphi}\hat{h}^{\mu\nu}\left[\bar{\chi}\left(i\gamma_\mu\overset{\leftrightarrow}{\partial}_\mu - i\eta_{\mu\nu}\overset{\leftrightarrow}{\slashed{\partial}}\right)\chi\right.\\
    &&~\left. \vphantom{\overset{\leftrightarrow}{\slashed{\partial}}}+ 8M_\chi e^A\eta_{\mu\nu}\bar{\chi}\chi\right]\delta(z-\bar{z}).
\end{eqnarray}

\subsubsection{Brane vector boson}
\label{sec:couplingbvector}
The 3-point interactions are given by the Lagrangian
\begin{eqnarray}
    \mathcal{L}_{h\bar{V}\bar{V}} &=& \dfrac{\kappa}{2}\int_{z_1}^{z_2}dz\left[\left(\hat{h}_{\mu\nu}-\dfrac{1}{4}\eta_{\mu\nu}\hat{h}\right)\bar{F}^{\mu\rho}\bar{F}^\nu{}_\rho \right.\\
    &&~\left. - m_{\bar{V}}^2\left(\hat{h}_{\mu\nu}-\dfrac{1}{2}\eta_{\mu\nu}\hat{h}\right)\bar{V}^\mu \bar{V}^\nu\right]\delta(z-\bar{z}),\\
    \mathcal{L}_{\varphi \bar{V}\bar{V}} &=& -\dfrac{\kappa}{2}\int_{z_1}^{z_2}dz\dfrac{m_{\bar{V}}^2}{\sqrt{6}}~\varphi\ \bar{V}_\mu \bar{V}^\mu~\delta(z-\bar{z}).
\end{eqnarray}
The 4-point interactions are given by the Lagrangian
\begin{eqnarray}
     \mathcal{L}_{hh\bar{V}\bar{V}} &=& \dfrac{\kappa^2}{4}\int_{z_1}^{z_2}dz ~\delta(z-\bar{z})\left\{\left[\vphantom{\dfrac{1}{2}}\hat{h}_{\mu\sigma}\hat{h}_{\nu\rho} + \eta_{\mu\rho}\left(\hat{h}\hat{h}_{\nu\sigma} - 2\hat{h}_{\nu}{}^{\alpha}\hat{h}_{\sigma\alpha}\right) \right.\right.\\
    &&~\left. + \dfrac{1}{4}\eta_{\mu\rho}\eta_{\nu\sigma}\left(\hat{h}^{\alpha\beta}\hat{h}_{\alpha\beta} - \dfrac{1}{2}\hat{h}^2\right)\right]\bar{F}^{\mu\nu}\bar{F}^{\rho\sigma} ,\\
    &&~ \left. - m_{\bar{V}}^2\left[\hat{h}\hat{h}_{\mu\nu} - 2\hat{h}_{\mu}{}^{\rho}\hat{h}_{\nu\rho} + \dfrac{1}{2}\eta_{\mu\nu}\left(\hat{h}^{\rho\sigma}\hat{h}_{\rho\sigma} - \dfrac{1}{2}\hat{h}^2\right)\right]\bar{V}^\mu \bar{V}^\nu\right\},\\
    \mathcal{L}_{\varphi\varphi \bar{V}\bar{V}} &=& \dfrac{\kappa^2}{4}\int_{z_1}^{z_2}dz~ \dfrac{m_{\bar{V}}^2}{6}\hat{\varphi}^2 \bar{V}^\mu \bar{V}_\mu~\delta(z-\bar{z}),\\
    \mathcal{L}_{h\varphi \bar{V}\bar{V}} &=& \dfrac{\kappa^2}{2}\int_{z_1}^{z_2}dz~ \dfrac{m_{\bar{V}}^2}{\sqrt{6}}\hat{\varphi}\left[ \left(\hat{h}_{\mu\nu}-\dfrac{1}{2}\eta_{\mu\nu}\hat{h}\right) \bar{V}^\mu \bar{V}^\nu\right]\delta(z-\bar{z}).
\end{eqnarray}

\subsection{Bulk Matter}
\subsubsection{Scalar}
\label{sec:couplingbulkscalar}
The 3-point interactions are given by the Lagrangian
\begin{eqnarray}
    \mathcal{L}_{hSS} &=& \dfrac{\kappa}{2}\int_{z_1}^{z_2}dz~e^{3A} \hat{h}^{\mu\nu}\left[-\partial_\mu S\partial_\nu S+\dfrac{1}{2}\eta_{\mu\nu}\left(\partial_\rho S\partial^\rho S - (\partial_z S)^2 -M_{S}^2e^{2A}S^2\right)\right],\\
    \mathcal{L}_{\varphi SS} &=& \dfrac{\kappa}{2}\int_{z_1}^{z_2}dz~ e^{3A}\frac{1}{\sqrt{6}}\hat{\varphi}\left[3\left(\partial_zS\right)^2+M_S^2e^{2A}S^2\right].
\end{eqnarray}
The 4-point interactions are given by the Lagrangian
\begin{eqnarray}
    \mathcal{L}_{hhSS} &=& \dfrac{\kappa^2}{4}\int_{z_1}^{z_2}dz~e^{3A} \left\{\vphantom{\frac{1}{4}}\left(2\hat{h}^{\mu\rho}\hat{h}^\nu{}_\rho-\hat{h}\hat{h}^{\mu\nu}\right)\partial_\mu S\partial_\nu S\right.\nonumber\\
    &&~\left.+\dfrac{1}{4}\left(\hat{h}^2-2\hat{h}^{\nu\rho}\hat{h}_{\nu\rho}\right)\left[\partial_\rho S\partial^\rho S - \left(\partial_z S\right)^2 -M_{S}^2e^{2A}S^2\right]\right\},\\
    \mathcal{L}_{\varphi\varphi SS} &=& -\dfrac{\kappa^2}{4}\int_{z_1}^{z_2}dz~e^{3A} \dfrac{1}{6}\hat{\varphi}^2\left[\partial^\mu S\partial_\mu S+10\left(\partial_z S\right)^2\right],\\
    \mathcal{L}_{h\varphi SS} &=& \dfrac{\kappa^2}{2}\int_{z_1}^{z_2}dz~e^{3A} \dfrac{1}{2\sqrt{6}}\hat{\varphi}\hat{h}\left[3\left(\partial_z S\right)^2 + M_S^2e^{2A}S^2\right].
\end{eqnarray}

\subsubsection{ Fermion}
\label{sec:couplingbulkfermion}
The 3-point interactions are given by the Lagrangian
\begin{eqnarray}
    \mathcal{L}_{h\bar{\psi}\psi} &=& \kappa\int_{z_1}^{z_2}dz~e^{4A} \left\{\dfrac{1}{4}\hat{h}^{\mu\nu}\left[\bar{\psi}_L\left(-i\gamma_\mu\overset{\leftrightarrow}{\partial}_\nu + i\eta_{\mu\nu}\overset{\leftrightarrow}{\slashed{\partial}}\right)\psi_L + (L\rightarrow R)\right]\right.\\
    &&~\left.-~\dfrac{1}{4}\hat{h}\left[\bar{\psi}_R\overset{\leftrightarrow}{\partial}_z\psi_L - (L\rightarrow R)\right] -\dfrac{1}{2}M_\psi e^A\hat{h}\left[\bar{\psi}_R\psi_L + \bar{\psi}_L\psi_R\right]\right\},\\
    \mathcal{L}_{\varphi\bar{\psi}\psi} &=& \kappa\int_{z_1}^{z_2}dz~e^{4A} \left\{\dfrac{1}{4\sqrt{6}}\hat{\varphi}\left[-i\bar{\psi}_L\overset{\leftrightarrow}{\slashed{\partial}}\psi_L + (L\rightarrow R)\right]\right.\\
    &&~\left.+\dfrac{1}{\sqrt{6}}\hat{\varphi}\left[\bar{\psi}_R\overset{\leftrightarrow}{\partial}_z\psi_L - (L\rightarrow R)\right] +~\dfrac{1}{\sqrt{6}}M_\psi e^A\hat{\varphi}\left[\bar{\psi}_R\psi_L + \bar{\psi}_L\psi_R\right]\right\},
\end{eqnarray}
where the derivative $\overset{\leftrightarrow}{\partial}$ acts only on the fermion fields and is defined as
\begin{equation}
    \overset{\leftrightarrow}{\partial}_M = \overset{\rightarrow}{\partial}_M - \overset{\leftarrow}{\partial}_M.
\end{equation}
The 4-point interactions are given by the Lagrangian
\begin{eqnarray}
    \mathcal{L}_{hh\bar{\psi}\psi} &=& \dfrac{\kappa^2}{2}\int_{z_1}^{z_2}dz~e^{4A} \left\{\dfrac{1}{8}\left(3\hat{h}^{\mu\rho}\hat{h}^\nu{}_\rho-2\hat{h}\hat{h}^{\mu\nu}\right)\left(i\bar{\psi}_L\gamma_\mu\overset{\leftrightarrow}{\partial}_\nu\psi_L + (L\leftrightarrow R)\right)\right.\nonumber\\
    &&~+\dfrac{1}{8}\left(\hat{h}^2-2\hat{h}^{\nu\rho}\hat{h}_{\nu\rho}\right)\left[i\bar{\psi}_L\overset{\leftrightarrow}{\slashed{\partial}}\psi_L - 2M_\psi e^A\bar{\psi}_R\psi_L+ (L\leftrightarrow R)\right]\\
    &&~\left. + \dfrac{1}{8}\epsilon^{\lambda\alpha\beta\rho}h^\mu{}_\alpha\partial_\rho h_{\mu\beta}\left[\bar{\psi}_L\gamma_\lambda \psi_L-(L\leftrightarrow R)\right]\right.\\
    &&~-\dfrac{1}{8}\left(\hat{h}^2-2\hat{h}^{\nu\rho}\hat{h}_{\nu\rho}\right)\left[\bar{\psi}_R\overset{\leftrightarrow}{\partial}_z\psi_L - (L\leftrightarrow R)\right]\\
    &&~\left.+\dfrac{1}{4}\hat{h}^{\mu\rho}\partial_z\hat{h}^\nu{}_\rho\left[ \bar{\psi}_R\sigma_{\mu\nu}\psi_L - (L\leftrightarrow R)\right]\right\}\\
    \mathcal{L}_{\varphi\varphi \bar{\psi}\psi} &=& -\dfrac{\kappa^2}{2}\int_{z_1}^{z_2}dz~e^{4A} \hat{\varphi}^2\left\{\left[\dfrac{i}{16}\bar{\psi}_L\overset{\leftrightarrow}{\slashed{\partial}}\psi_L + (L\leftrightarrow R)\right] \right.\\
    &&~\left.+ \left[\dfrac{1}{3}\bar{\psi}_R\overset{\leftrightarrow}{\partial}_z\psi_L - (L\leftrightarrow R)\right]\right\},\\
    \mathcal{L}_{h\varphi \bar{\psi}\psi} &=& \kappa^2\int_{z_1}^{z_2}dz~e^{3A} \dfrac{1}{8\sqrt{6}}\hat{\varphi}\hat{h}^{\mu\nu}\left\{\left[\bar{\psi}_L\left(i\gamma_\mu\overset{\leftrightarrow}{\partial}_\mu - i\eta_{\mu\nu}\overset{\leftrightarrow}{\slashed{\partial}}\right)\psi_L \right.\right.\\
    &&~\left. \vphantom{\overset{\leftrightarrow}{\slashed{\partial}}}+ 4M_\psi e^A\eta_{\mu\nu}\bar{\psi}_R\psi_L + (L\leftrightarrow R)\right] \\
    &&~ + \left.\left[4\eta_{\mu\nu}\bar{\psi}_R\overset{\leftrightarrow}{\partial}_z\psi_L - (L\leftrightarrow R)\right]\right\}.
\end{eqnarray}

\subsubsection{Vector boson}
\label{sec:couplingbulkvector}

The 3-point interactions are given by the Lagrangian
\begin{eqnarray}
    \mathcal{L}_{hVV} &=& \dfrac{\kappa}{2}\int_{z_1}^{z_2}dz~e^{A} \left[\left(\hat{h}_{\mu\nu}-\dfrac{1}{4}\eta_{\mu\nu}\hat{h}\right)F^{\mu\rho}F^\nu{}_\rho \right.\\
    &&~ \left. - \left(\hat{h}_{\mu\nu}-\dfrac{1}{2}\eta_{\mu\nu}\hat{h}\right)\partial_z V^\mu\partial_z V^\nu\right],\\
    \mathcal{L}_{hV_5V_5} &=& -\dfrac{\kappa}{2}\int_{z_1}^{z_2}dz~e^{A}  \left(\hat{h}_{\mu\nu}-\dfrac{1}{2}\eta_{\mu\nu}\hat{h}\right)\partial_\mu V_5\partial_\nu V_5,\\
    \mathcal{L}_{hVV_5} &=& \kappa\int_{z_1}^{z_2}dz~e^{A}  \left(\hat{h}_{\mu\nu}-\dfrac{1}{2}\eta_{\mu\nu}\hat{h}\right)\partial_\mu V_5\partial_z V_\nu,\\
    \mathcal{L}_{\varphi VV} &=& \dfrac{\kappa}{2}\int_{z_1}^{z_2}dz~e^{A}\sqrt{\dfrac{2}{3}}~\varphi\ \left(-\dfrac{1}{4}F^{\mu\nu}F_{\mu\nu} - \partial_zV^\mu\partial_zV_\mu\right),\\
    \mathcal{L}_{\varphi V_5V_5} &=& -\dfrac{\kappa}{2}\int_{z_1}^{z_2}dz~e^{A}\sqrt{\dfrac{2}{3}}~\varphi\ \partial_\mu V_5\partial^\mu V_5,\\
    \mathcal{L}_{\varphi VV_5} &=& \kappa\int_{z_1}^{z_2}dz~e^{A}\sqrt{\dfrac{2}{3}}~\varphi\ \partial_\mu V_5\partial_z V^\mu.
\end{eqnarray}
The 4-point interactions are given by the Lagrangian
\begin{eqnarray}
    \mathcal{L}_{hhVV} &=& \dfrac{\kappa^2}{4}\int_{z_1}^{z_2}dz~e^{A} \left\{\left[\vphantom{\dfrac{1}{2}}\hat{h}_{\mu\sigma}\hat{h}_{\nu\rho} + \eta_{\mu\rho}\left(\hat{h}\hat{h}_{\nu\sigma} - 2\hat{h}_{\nu}{}^{\alpha}\hat{h}_{\sigma\alpha}\right) \right.\right.\\
    &&~\left.\left. + \dfrac{1}{4}\eta_{\mu\rho}\eta_{\nu\sigma}\left(\hat{h}^{\alpha\beta}\hat{h}_{\alpha\beta} - \dfrac{1}{2}\hat{h}^2\right)\right]F^{\mu\nu}F^{\rho\sigma} \right.\\
    &&~ \left. - \left[\hat{h}\hat{h}_{\mu\nu} - 2\hat{h}_{\mu}{}^{\rho}\hat{h}_{\nu\rho} + \dfrac{1}{2}\eta_{\mu\nu}\left(\hat{h}^{\rho\sigma}\hat{h}_{\rho\sigma} - \dfrac{1}{2}\hat{h}^2\right)\right]\partial_z V^\mu\partial_z V^\nu\right\},\\
    \mathcal{L}_{\varphi\varphi VV} &=& \dfrac{\kappa^2}{4}\int_{z_1}^{z_2}dz~e^{A} \dfrac{5}{6}\hat{\varphi}^2 \partial_zV^\mu\partial_zV_\mu,\\
    \mathcal{L}_{h\varphi VV} &=& \dfrac{\kappa^2}{2}\int_{z_1}^{z_2}dz~e^{A} \dfrac{1}{\sqrt{6}}\hat{\varphi}\left[\left(\hat{h}_{\mu\nu}-\dfrac{1}{4}\eta_{\mu\nu}\hat{h}\right)F^{\mu\rho}F^\nu{}_\rho \right.\\
    &&~\left. + \left(2\hat{h}_{\mu\nu}-\eta_{\mu\nu}\hat{h}\right)\partial_z V^\mu\partial_z V^\nu\right].
\end{eqnarray}

\section{Wavefunctions of bulk matter}
\label{app:wfnbulk}
\subsection{Graviton}
\label{sec:wfngraviton}
The gravitational wavefunctions in RS, in conformal coordinates, take the form of
\begin{eqnarray}
    f^{(n)}(z) &=& C^{(n)}_h z^2\left[Y_1(m_n z_2)J_2(m_n z)-J_1(m_n z_2)Y_2(m_n z)\right], \\
    g^{(n)}(z) &=& C^{(n)}_A z^2\left[Y_1(m_n z_2)J_1(m_n z)-J_1(m_n z_2)Y_1(m_n z)\right], \\
    k^{(n)}(z) &=& C^{(n)}_\varphi z^2\left[Y_1(m_n z_2)J_0(m_n z)-J_1(m_n z_2)Y_0(m_n z)\right],
\end{eqnarray}
for the massive modes $n>0$, where $J_{a}$ and $Y_{a}$ are Bessel functions of the first and second kind, respectively, and
\begin{eqnarray}
    f^{(0)}(z) &=& C^{(0)}_h, \\
    g^{(0)}(z) &=& 0, \\
    k^{(0)}(z) &=& C^{(0)}_\varphi z^2,
\end{eqnarray}
for the massless modes. The normalizations $C^{(n)}_{h,A,\varphi}$ are fixed by
\begin{equation}
\aligned
    &\int_{z_1}^{z_2}dz~e^{3A(z)}f^{(m)}(z)f^{(n)}(z) = \int_{z_1}^{z_2}dz~e^{3A(z)}g^{(m)}(z)g^{(n)}(z)\\
    =~& \int_{z_1}^{z_2}dz~e^{3A(z)}k^{(m)}(z)k^{(n)}(z) = \delta_{m,n}.
\endaligned
\end{equation}
The physical mass $m_n$ is the $n$-th solution of the equation
\begin{equation}
    Y_1(m_n z_2)J_1(m_n z_1)-J_1(m_n z_2)Y_1(m_n z_1) = 0.
\end{equation}

\subsection{Bulk scalar}
\label{sec:wfnscalar}

The wavefunctions of KK scalars are given by
\begin{equation}
    f_S^{(n)}(z) = z^2\left[c_n Y_{\nu}(m_{S,n} z) + d_n J_{\nu}(m_{S,n} z)\right],
\end{equation}
where $\nu=\sqrt{4+M_S^2 z_1^2}$, and the coefficents $c_n$ and $d_n$ and the masses $m_{S,n}$ are fixed by the boundary conditions,
\begin{equation}
    \partial_z f^{(n)}_S(z_1) = \partial_z f^{(n)}_S(z_2) = 0,
\end{equation}and orthogonality,
\begin{equation}
    \int_{z_1}^{z_2}dz~ e^{3A(z)} f_{S}^{(m)}(z)f_{S}^{(n)}(z) = \delta_{m,n}.
\end{equation}

\subsection{Bulk fermion}
\label{sec:wfnfermion}
Without the loss of generality, we consider the case the left-handed fermion has a massless mode. In such case, the wavefunctions are given by
\begin{eqnarray}
    f_{\psi_L}^{(0)}(z) &=& C^{(0)}_{\psi_L} z^{2-M_\psi  z_1},\\
    f_{\psi_R}^{(0)}(z) &=& 0,\\
    f_{\psi_L}^{(n)}(z) &=& C^{(n)}_{\psi_L} z^{\frac{5}{2}}\left(Y_{M_\psi z_1+1/2}(m_{\psi,n} z) - \dfrac{Y_{M_\psi  z_1-1/2}(m_{\psi,n} z_2)J_{M_\psi  z_1+1/2}(m_{\psi,n} z)}{J_{M_\psi  z_1-1/2}(m_{\psi,n} z_2)}\right),\\
    f_{\psi_R}^{(n)}(z) &=& C^{(n)}_{\psi_R} z^{\frac{5}{2}}\left(Y_{M_\psi  z_1-1/2}(m_{\psi,n} z) - \dfrac{Y_{M_\psi  z_1-1/2}(m_{\psi,n} z_2)J_{M_\psi  z_1-1/2}(m_{\psi,n} z)}{J_{M_\psi  z_1-1/2}(m_{\psi,n} z_2)}\right).
\end{eqnarray}
where the masses $m_{\psi,n}$ are the solutions of the equation
\begin{equation}
    J_{M_\psi  z_1-1/2}(m_{\psi,n} z_2)Y_{M_\psi  z_1-1/2}(m_{\psi,n} z_1) - Y_{M_\psi  z_1-1/2}(m_{\psi,n} z_2)J_{M_\psi  z_1-1/2}(m_{\psi,n} z_1) = 0,
\end{equation}
and the normalization $C^{(n)}_{\psi_{L/R}}$ are fixed by the orthogonality
\begin{equation}
    \int_{z_1}^{z_2}dz~ e^{4A(z)} f_{\psi_{L/R}}^{(m)}(z)f_{\psi_{L/R}}^{(n)}(z) = \delta_{m,n}.
\end{equation}

\subsection{Bulk vector}
\label{sec:wfnbulkvector}
The wavefunctions of KK gauge bosons are given by,
\begin{eqnarray}
    f_V^{(0)}(z) &=& C^{(0)}_{V},\\
    f_{V_5}^{(0)}(z) &=& 0,\\
    f_V^{(n)}(z) &=& C^{(n)}_{V} z\left(J_{1}(m_{V,n} z) - \dfrac{J_{0}(m_{V,n} z_1)Y_{1}(m_{V,n} z)}{Y_{0}(m_{V,n} z_1)}\right),\\
    f_{V_5}^{(n)}(z) &=& C^{(n)}_{V_5} z\left(J_{0}(m_{V,n} z) - \dfrac{J_{0}(m_{V,n} z_1)Y_{0}(m_{V,n} z)}{Y_{0}(m_{V,n} z_1)}\right),
\end{eqnarray}
where the masses $m_{V,n}$ are the solutions of the equation
\begin{equation}
    Y_0(m_{V,n}z_1)J_0(m_{V,n}z_2) - J_0(m_{V,n}z_1)Y_0(m_{V,n}z_2) = 0,
\end{equation}
and the normalization $C^{(n)}_V$ and $C^{(n)}_{V_5}$ are fixed by the orthogonality
\begin{equation}
    \int_{z_1}^{z_2}dz~ e^{A(z)} f_{V}^{(m)}(z)f_{V}^{(n)}(z) = \int_{z_1}^{z_2}dz~ e^{A(z)} f_{V_5}^{(m)}(z)f_{V_5}^{(n)}(z) = \delta_{m,n}.
\end{equation}

\section{Coupling Structures}
\label{app:coup}
\subsection{Graviton}
The overlap integrals relevant to the KK graviton self interaction are given by
\begin{equation}
    a_{n_1n_2n_3} = \braket{f^{(n_1)}f^{(n_2)}f^{(n_3)}},\quad
    b_{\bar{n}_1\bar{n}_2n_3} = \braket{(\partial_zf^{(n_1)})(\partial_zf^{(n_1)})f^{(n_3)}},\quad
    b_{\bar{n}_1\bar{n}_2r} = \braket{(\partial_zf^{(n_1)})(\partial_zf^{(n_1)})k^{(0)}}.
\end{equation}
One can derive the following "$b$-to-$a$" identities using the eigenequations and integration by parts,
\begin{eqnarray}
    &&b_{\bar{j}\bar{n}n} + b_{\bar{n}\bar{n}j} = m_n^2a_{nnj},\\
    &&b_{\bar{n}\bar{n}j} = \left(m_n^2 - \frac{1}{2}m_j^2\right)a_{nnj}.
\end{eqnarray}

\subsection{Bulk scalar}
\label{sec:overlapscalar}
The overlap integrals relevant to the KK scalars are given by
{\belowdisplayskip=0pt
\begin{alignat}{6}
    &a^{S}_{n_1n_2n_3} &=& \braket{f^{(n_1)}f_{S}^{(n_2)}f_{S}^{(n_3)}}_S,\quad
    &&a^{M_S}_{n_1n_2n_3} &=& \braket{e^{2A}f^{(n_1)}f_{S}^{(n_2)}f_{S}^{(n_3)}}_S,\quad
    &&b^{S}_{n_1\bar{n}_2\bar{n}_3} &=&\braket{f^{(n_1)}\partial_zf_{S}^{(n_2)}\partial_zf_{S}^{(n_3)}}_S,\\
    &a^{S}_{n_1n_2r} &=& \braket{f_{S}^{(n_1)}f_{S}^{(n_2)}k^{(0)}}_S,\quad
    &&a^{M_S}_{n_1n_2r} &=& \braket{e^{2A}f_{S}^{(n_1)}f_{S}^{(n_2)}k^{(0)}}_S,\quad
    &&b^{S}_{\bar{n}_1\bar{n}_2r} &=& \braket{\partial_zf_{S}^{(n_1)}\partial_zf_{S}^{(n_2)}k^{(0)}}_S,
\end{alignat}}
{
\abovedisplayskip=0pt
\begin{alignat}{4}
    &a^{S}_{n_1n_2n_3n_4} &=& \braket{f^{(n_1)}f^{(n_2)}f_{S}^{(n_3)}f_{S}^{(n_4)}}_S,\quad
    &&a^{M_S}_{n_1n_2n_3n_4} &=& \braket{e^{2A}f^{(n_1)}f^{(n_2)}f_{S}^{(n_3)}f_{S}^{(n_4)}}_S,\\
    &b^{S}_{\bar{n}_1\bar{n}_2n_3n_4} &=& \braket{\partial_zf^{(n_1)}\partial_zf^{(n_2)}f_{S}^{(n_3)}f_{S}^{(n_4)}}_S,\quad
    &&b^{S}_{n_1n_2\bar{n}_3\bar{n}_4} &=& \braket{f^{(n_1)}f^{(n_2)}\partial_zf_S^{(n_3)}\partial_zf_{S}^{(n_4)}}_S,\\
    &b^{M_S}_{\bar{n}_1\bar{n}_2n_3n_4} &=& \braket{e^{2A}\partial_zf^{(n_1)}\partial_zf^{(n_2)}f_{S}^{(n_3)}f_{S}^{(n_4)}}_S.
\end{alignat}}

One can derive the following "$b$-to-$a$" identities using the eigenequations and integration by parts,
\begin{eqnarray}
    && b^S_{j\bar{m}\bar{m}} = \left(m_{S,m}^2 - \frac{1}{2}m_{j}^2\right)a^S_{jmm} - M_S^2a^{M_S}_{jmm},\\
    && b^S_{nn\bar{m}\bar{m}} = b^S_{\bar{n}\bar{n}mm} + \left(m_{S,m}^2-m_n^2\right)a^S_{nnmm} - M_S^2a^{M_S}_{nnmm}.
\end{eqnarray}

The completeness relations can be written as
\begin{equation}
    \sum_{j=0}^{\infty} \left(a^{S}_{nmj}\right)^2 = a^{S}_{nnmm}, \qquad
    \sum_{j=0}^{\infty} a_{nnj}a^{S}_{jmm} = a^{S}_{nnmm}, \qquad
    \sum_{j=0}^{\infty} b_{\bar{n}\bar{n}j}a^{S}_{jmm} = b^{S}_{\bar{n}\bar{n}mm}.
\end{equation}

\subsection{Bulk fermions}
\label{sec:overlapfermion}
The overlap integrals relevant to the KK fermions are given by
{\belowdisplayskip=0pt
\begin{alignat}{6}
    &a^{\psi_{L/R}}_{n_1n_2n_3} &=& \braket{f^{(n_1)}f_{\psi_{L/R}}^{(n_2)}f_{\psi_{L/R}}^{(n_3)}}_\psi,\quad
    &&a^{M_\psi}_{n_1n_2n_3} &=& \braket{e^{A}f^{(n_1)}f_{\psi_L}^{(n_2)}f_{\psi_R}^{(n_3)}}_\psi,\quad
    &&b^{\psi_L\psi_R}_{\bar{n}_1n_2n_3} &=& \braket{(\partial_zf^{(n_1)})f_{\psi_L}^{(n_2)}f_{\psi_R}^{(n_3)}}_\psi,\\
    &b^{\psi_L\psi_R}_{n_1\bar{n}_2n_3} &=& \braket{f^{(n_1)}(\partial_zf_{\psi_L}^{(n_2)})f_{\psi_R}^{(n_3)}}_\psi,\quad
    &&b^{\psi_L\psi_R}_{n_1n_2\bar{n}_3} &=& \braket{f^{(n_1)}f_{\psi_L}^{(n_2)}(\partial_zf_{\psi_R}^{(n_3)})}_\psi,\quad
    &&a^{\psi_{L/R}}_{n_1n_2r} &=& \braket{f_{\psi_{L/R}}^{(n_1)}f_{\psi_{L/R}}^{(n_2)}k^{(0)}}_\psi,\\
    &a^{M_\psi}_{n_1n_2r} &=& \braket{e^{A}f_{\psi_L}^{(n_1)}f_{\psi_R}^{(n_2)}k^{(0)}}_\psi,\quad
    &&b^{\psi_L\psi_R}_{\bar{n}_1n_2r} &=& \braket{(\partial_zf_{\psi_L}^{(n_1)})f_{\psi_R}^{(n_2)}k^{(0)}}_\psi,\quad
    &&b^{\psi_L\psi_R}_{n_1\bar{n}_2r} &=& \braket{f_{\psi_L}^{(n_1)}(\partial_zf_{\psi_R}^{(n_2)})k^{(0)}}_\psi,
\end{alignat}}
{\abovedisplayskip=0pt
\begin{alignat}{4}
    &a^{\psi_{L/R}}_{n_1n_2n_3n_4} &=& \braket{f^{(n_1)}f^{(n_2)}f_{\psi_{L/R}}^{(n_3)}f_{\psi_{L/R}}^{(n_4)}}_\psi,\quad
    &&a^{M_\psi}_{n_1n_2n_3n_4} &=& \braket{e^Af^{(n_1)}f^{(n_2)}f_{\psi_L}^{(n_3)}f_{\psi_R}^{(n_4)}}_\psi,\\
    &b^{\psi_L\psi_R}_{\bar{n}_1n_2n_3n_4} &=& \braket{(\partial_zf^{(n_1)})f^{(n_2)}f_{\psi_L}^{(n_3)}f_{\psi_R}^{(n_4)}}_\psi,\quad
    &&b^{\psi_L\psi_R}_{n_1n_2\bar{n}_3n_4} &=& \braket{f^{(n_1)}f^{(n_2)}(\partial_zf_{\psi_L}^{(n_3)})f_{\psi_R}^{(n_4)}}_\psi,\\
    &b^{\psi_L\psi_R}_{n_1n_2n_3\bar{n}_4} &=& \braket{f^{(n_1)}f^{(n_2)}f_{\psi_L}^{(n_3)}(\partial_zf_{\psi_R}^{(n_4)})}_\psi,\quad
    &&b^{\psi_L}_{\bar{n}_1\bar{n}_2n_3n_4} &=& \braket{(\partial_zf^{(n_1)})(\partial_zf^{(n_2)})f_{\psi_L}^{(n_3)}f_{\psi_L}^{(n_4)}}_\psi,\\
    &b^{\psi_R}_{\bar{n}_1\bar{n}_2n_3n_4} &=& \braket{(\partial_zf^{(n_1)})(\partial_zf^{(n_2)})f_{\psi_R}^{(n_3)}f_{\psi_R}^{(n_4)}}_\psi,\quad
    &&b^{M_\psi}_{\bar{n}_1\bar{n}_2n_3n_4} &=& \braket{e^A(\partial_zf^{(n_1)})(\partial_zf^{(n_2)})f_{\psi_L}^{(n_3)}f_{\psi_R}^{(n_4)}}_\psi.
\end{alignat}}
One can derive the following "$b$-to-$a$" identities using the eigenequations and integration by parts,
\begin{eqnarray}
    && b^{\psi_L\psi_R}_{\bar{n}mj} = m_{\psi,j}a^{\psi_L}_{nmj} - m_{\psi,m}a^{\psi_R}_{nmj}, \label{eq:proof_ref_1}\\
    && b^{\psi_L\psi_R}_{\bar{n}jm} = -m_{\psi,j}a^{\psi_R}_{nmj} + m_{\psi,m}a^{\psi_L}_{nmj},\\
    &&b^{\psi_L\psi_R}_{jm\bar{m}} = b^{\psi_L\psi_R}_{j\bar{m}m} - m_{\psi,m}a^{\psi_L}_{jmm} - m_{\psi,m}a^{\psi_R}_{jmm} + 2M_\psi a^{M_\psi}_{jmm},\\
    &&b^{\psi_L\psi_R}_{\bar{m}mr} = m_{\psi,m}a^{\psi_{L}}_{mmr}-M_\psi a^{M_\psi}_{mmr},\\
    &&b^{\psi_L\psi_R}_{m\bar{m}r} = -m_{\psi,m}a^{\psi_{R}}_{mmr} + M_\psi a^{M_\psi}_{mmr},\\
    &&b^{\psi_L\psi_R}_{\bar{n}nmm}=\frac{1}{2}m_{\psi,m}a^{\psi_L}_{nnmm}-\frac{1}{2}m_{\psi,m}a^{\psi_R}_{nnmm},\label{eq:proof_ref_3}\\
    &&b^{\psi_L\psi_R}_{nnm\bar{m}} = b^{\psi_L\psi_R}_{nn\bar{m}m} - m_{\psi,m}a^{\psi_L}_{nnmm} - m_{\psi,m}a^{\psi_R}_{nnmm} + 2M_\psi a^{M_\psi}_{nnmm}.
\end{eqnarray}

The completeness relations can be written as
\begin{alignat}{6}
    \sum_{j=0}^{\infty} \left(a^{\psi_{L/R}}_{nmj}\right)^2 &=&& a^{\psi_{L/R}}_{nnmm}, \qquad
    &\sum_{j=0}^{\infty} a^{\psi_{L}}_{nmj}b^{\psi_L\psi_R}_{\bar{n}jm} &=&& b^{\psi_L\psi_R}_{\bar{n}nmm},\qquad
    &\sum_{j=0}^{\infty} a^{\psi_{R}}_{nmj}b^{\psi_L\psi_R}_{\bar{n}mj} &=&& b^{\psi_L\psi_R}_{\bar{n}nmm}, \label{eq:proof_ref_2}\\
    \sum_{j=0}^{\infty} a_{nnj}a^{\psi_{L/R}}_{jmm} &=&& a^{\psi_{L/R}}_{nnmm},\qquad
    &\sum_{j=0}^{\infty} b_{\bar{n}\bar{n}j}a^{\psi_{L/R}}_{jmm} &=&& b^{\psi_{L/R}}_{\bar{n}\bar{n}mm},\qquad
    &\sum_{j=0}^{\infty} \left(b^{\psi_L\psi_R}_{\bar{n}jm}\right)^2 &=&& b^{\psi_{R}}_{\bar{n}\bar{n}mm}, \\
    \sum_{j=0}^{\infty} \left(b^{\psi_L\psi_R}_{\bar{n}mj}\right)^2 &=&& b^{\psi_{L}}_{\bar{n}\bar{n}mm}.
\end{alignat}

\subsection{Bulk gauge bosons}
\label{sec:overlapvector}
The overlap integrals relevant to the KK gauge boson are given by
{\belowdisplayskip=0pt
\begin{alignat}{6}
    &a^{V}_{n_1n_2n_3} &=& \braket{f^{(n_1)}f_{V}^{(n_2)}f_{V}^{(n_3)}}_V,\quad
    &&a^{V_5}_{n_1n_2n_3} &=& \braket{f^{(n_1)}f_{V_5}^{(n_2)}f_{V_5}^{(n_3)}}_V,\quad
    &&b^{V}_{n_1\bar{n}_2\bar{n}_3} &=& \braket{f^{(n_1)}(\partial_zf_{V}^{(n_2)})(\partial_zf_{V}^{(n_3)})}_V,\\
    &a^{V}_{n_1n_2r} &=& \braket{f_{V}^{(n_1)}f_{V}^{(n_2)}k^{(0)}}_V,\quad
    &&b^{V}_{\bar{n}_1\bar{n}_2r} &=& \braket{(\partial_zf_{V}^{(n_1)})(\partial_zf_{V}^{(n_2)})k^{(0)}}_V,
\end{alignat}}
{\abovedisplayskip=0pt
\begin{alignat}{4}
    &a^{V}_{n_1n_2n_3n_4} &=& \braket{f^{(n_1)}f^{(n_2)}f_{V}^{(n_3)}f_{V}^{(n_4)}}_V,\quad
    &&a^{V_5}_{n_1n_2n_3n_4} &=& \braket{f^{(n_1)}f^{(n_2)}f_{V_5}^{(n_3)}f_{V_5}^{(n_4)}}_V,\\
    &b^{V}_{\bar{n}_1\bar{n}_2n_3n_4} &=& \braket{(\partial_zf^{(n_1)})(\partial_zf^{(n_2)})f_{V}^{(n_3)}f_{V}^{(n_4)}}_V,\quad
    &&b^{V}_{n_1n_2\bar{n}_3\bar{n}_4} &=& \braket{f^{(n_1)}f^{(n_2)}(\partial_zf_V^{(n_3)})(\partial_zf_V^{(n_4)})}_V,\\
    &b^{V_5}_{\bar{n}_1\bar{n}_2n_3n_4} &=& \braket{(\partial_zf^{(n_1)})(\partial_zf^{(n_2)})f_{V_5}^{(n_3)}f_{V_5}^{(n_4)}}_V.
\end{alignat}}

One can derive the following "$b$-to-$a$" identities using the eigenequations and integration by parts,
\begin{eqnarray}
    && b^V_{n_1\bar{n}_2\bar{n}_3} = m_{V,n_2}m_{V,n_3}a^{V_5}_{n_1n_2n_3},\\
    && b^V_{n_1n_2\bar{n}_3\bar{n}_4} = m_{V,n_3}m_{V,n_4}a^{V_5}_{n_1n_2n_3n_4}.
\end{eqnarray}

The completeness relations can be written as
\begin{alignat}{6}
    \sum_{j=0}^{\infty} \left(a^{V}_{nmj}\right)^2 &=&& a^{V}_{nnmm}, \qquad
    \sum_{j=0}^{\infty} \left(a^{V_5}_{nmj}\right)^2 &=&& a^{V_5}_{nnmm}, \qquad
    \sum_{j=0}^{\infty} a_{nnj}a^{V}_{jmm} &=&& a^{V}_{nnmm},\\
    \sum_{j=0}^{\infty} a_{nnj}a^{V_5}_{jmm} &=&& a^{V_5}_{nnmm},\qquad
    \sum_{j=0}^{\infty} b_{\bar{n}\bar{n}j}a^{V}_{jmm} &=&& b^{V}_{\bar{n}\bar{n}mm},\qquad
    \sum_{j=0}^{\infty} b_{\bar{n}\bar{n}j}a^{V_5}_{jmm} &=&& b^{V_5}_{\bar{n}\bar{n}mm}.
\end{alignat}

\section{Kinematics}
\label{sec:kinematics}

We define the Mandelstam variables such that,
\begin{align}
    s & = \left(p_{1} + p_{2}\right)^{2} = \left(k_{1} + k_{2}\right)^{2}, \\
    t & = \left(p_{1} - k_{1}\right)^{2} = \left(p_{2} - k_{2}\right)^{2}, \\
    u & = \left(p_{1} - k_{2}\right)^{2} = \left(p_{2} - k_{1}\right)^{2}.
\end{align}
Choosing the $\hat{z}$ direction as the centre-of-momentum frame, with two outgoing massive spin-2 KK graviton with masses $m_{n}$, and two incoming massive particles with masses $m_{m}$, we can express the four-momenta of various particles as,
\begin{eqnarray}
    & p_{1}^{\mu} = \dfrac{\sqrt{s}}{2}\left(1,0,0,\beta_{i,m}\right),\quad  & p_{2}^{\mu} = \dfrac{\sqrt{s}}{2}\left(1,0, 0, -\beta_{i,m}\right),\\
    & k_{1}^{\mu} = \dfrac{\sqrt{s}}{2}\left(1, \beta_{n}\sin\theta, 0, \beta_{n}\cos\theta\right),\quad & k_{2}^{\mu} = \dfrac{\sqrt{s}}{2}\left(1, -\beta_{n}\sin\theta, 0, -\beta_{n}\cos\theta\right),
\end{eqnarray}
where $\beta_{n} = \sqrt{1-4m_{n}^{2}/s}$, and $\beta_{i, m} = \sqrt{1-4m_{i,m}^2/s}$ for $i = \bar{S},S^{(m)},\chi,\psi^{(m)},\bar{V},V^{(m)}$.

\section{Proof of sum-rules}
\label{sec:sumrules}

In this section, we give the analytic proof of the sum-rules for bulk fermions. The proof for bulk scalars and gauge bosons can be easily derived in a similar manner.

\begin{itemize}
    \item Sum-rule at $\mathcal{O}(s^{5/2})$, given in Eq.~(\ref{sr:fermion_2}).
    \begin{equation}
    \aligned
        2\sum_{j=0}^{\infty}m_{\psi,j} a^{\psi_L}_{nmj}a^{\psi_R}_{nmj} =~& 2\sum_{j=0}^{\infty}\left(b^{\psi_L\psi_R}_{\bar{n}mj} + m_{\psi,m}a^{\psi_R}_{nmj}\right)a^{\psi_R}_{nmj} \\
        =~& 2b^{\psi_L\psi_R}_{\bar{n}nmm} + 2m_{\psi,m} a^{\psi_R}_{nnmm} \\
        =~& m_{\psi,m}a^{\psi_L}_{nnmm}+m_{\psi,m}a^{\psi_R}_{nnmm},
    \endaligned
    \end{equation}
    where we have used Eqs.~(\ref{eq:proof_ref_1}), (\ref{eq:proof_ref_2}) and (\ref{eq:proof_ref_3}).
    \item Sum-rules at $\mathcal{O}(s^{2})$, given in Eq.~(\ref{sr:fermion_2}) and (\ref{sr:fermion_4}).

    With the "$b$-to-$a$" relations and the completeness relations given in App.~\ref{sec:overlapfermion}, one can derive,
    \begin{eqnarray}
        \sum_{j=0}^{\infty}m_{\psi,j}^2\left(a^{\psi_L}_{nmj}\right)^2
        &=~&\sum_{j=0}^{\infty}\left[m_{\psi,j}a^{\psi_L}_{nmj}-m_{\psi,m}a^{\psi_R}_{nmj}\right]^2 - \sum_{j=0}^{\infty}m_{\psi,m}^2\left(a^{\psi_R}_{nmj}\right)^2 \nonumber\\
        &&~+ 2\sum_{j=0}^{\infty}m_{\psi,m}m_{\psi,j}a^{\psi_L}_{nmj}a^{\psi_R}_{nmj} \\
        &=~&\sum_{j=0}^{\infty}\left(b^{\psi_L\psi_R}_{\bar{n}mj}\right)^2 + m_{\psi,m}^2a^{\psi_L}_{nnmm} \\
        &=~&b^{\psi_L}_{\bar{n}\bar{n}mm} + m_{\psi,m}^2a^{\psi_L}_{nnmm},
    \end{eqnarray}
    \begin{eqnarray}
        \sum_{j=0}^{\infty} m_j^2 a_{nnj}a^{\psi_L}_{jmm} &=&~ \sum_{j=0}^{\infty} (2m_n^2a_{nnj} - 2b_{\bar{n}\bar{n}j})a^{\psi_L}_{jmm} \\
        &=&~2m_n^2a^{\psi_L}_{nnmm}  - 2b^{\psi_L}_{\bar{n}\bar{n}mm}.
    \end{eqnarray}
    \item To prove the sum-rule at $\mathcal{O}(s^{3/2})$, as given in Eq.~(\ref{sr:fermion_5}), we first show
    \begin{equation}
        \sum_{j=0}^{\infty}m_{\psi,j}b^{\psi_L\psi_R}_{\bar{n}jm}b^{\psi_L\psi_R}_{\bar{n}mj} = - \frac{1}{2}m_{\psi,m}\left(b^{\psi_{L}}_{\bar{n}\bar{n}mm} + b^{\psi_{R}}_{\bar{n}\bar{n}mm}\right)+ 2M_\psi b^{M_\psi}_{\bar{n}\bar{n}mm}.
    \end{equation}
    {\bf Proof:} Note that, using the eigen-equations and integration by parts,
    \begin{equation}
    \aligned
        \frac{m_{\psi,j}}{m_n}b^{\psi_L\psi_R}_{\bar{n}jm} =~& \frac{m_{\psi,j}}{m_n}\braket{(\partial_z f^{(n)}) f^{(j)}_{\psi_L} f^{(m)}_{\psi_R}}_\psi\\
        =~& -\braket{g^{(n)} f^{(m)}_{\psi_R} (\partial_z+2A' - M_\psi e^A)f^{(j)}_{\psi_R}}_\psi \\
        =~& \braket{f^{(j)}_{\psi_R} (\partial_z+2A' + M_\psi e^A) (g^{(n)} f^{(m)}_{\psi_R}) }_\psi\\
        =~& -m_n\braket{f^{(n)}f^{(m)}_{\psi_R}f^{(j)}_{\psi_R}}_\psi - m_{\psi,m} \braket{g^{(n)}f^{(m)}_{\psi_L}f^{(j)}_{\psi_R}}_\psi\\
        &~ + 2M_\psi\braket{e^A g^{(n)}f^{(m)}_{\psi_R}f^{(j)}_{\psi_R}}_\psi - 3 \braket{A'g^{(n)}f^{(m)}_{\psi_R}f^{(j)}_{\psi_R}}_\psi.
    \endaligned
    \end{equation}
    Thus, with the completeness relations, 
    \begin{equation}
    \aligned
        \sum_{j=0}^{\infty}m_{\psi,j}b^{\psi_L\psi_R}_{\bar{n}jm}b^{\psi_L\psi_R}_{\bar{n}mj} =~& \sum_{j=0}^{\infty}\left( -m_n^3 \braket{f^{(n)}f^{(m)}_{\psi_R}f^{(j)}_{\psi_R}} \braket{g^{(n)}f^{(m)}_{\psi_L}f^{(j)}_{\psi_R}}_\psi\right.\\
        &~ - m_n^2m_{\psi,m} \braket{g^{(n)}f^{(m)}_{\psi_L}f^{(j)}_{\psi_R}} \braket{g^{(n)}f^{(m)}_{\psi_L}f^{(j)}_{\psi_R}}_\psi\\
        &~+ 2 m_n^2M_\psi\braket{e^Ag^{(n)}f^{(m)}_{\psi_R}f^{(j)}_{\psi_R}} \braket{g^{(n)}f^{(m)}_{\psi_L}f^{(j)}_{\psi_R}}_\psi\\
        &~ \left.- 3m_n^2\braket{A'g^{(n)}f^{(m)}_{\psi_R}f^{(j)}_{\psi_R}} \braket{g^{(n)}f^{(m)}_{\psi_L}f^{(j)}_{\psi_R}}_\psi\right) \\
        =~& -m_n^3 \braket{f^{(n)}g^{(n)}f^{(m)}_{\psi_L}f^{(m)}_{\psi_R}}_\psi - m_n^2m_{\psi,m}\braket{g^{(n)}g^{(n)}f^{(m)}_{\psi_L}f^{(m)}_{\psi_L}}_\psi \\
        & - 3m_n^2\braket{A'g^{(n)}g^{(n)}f^{(m)}_{\psi_L}f^{(m)}_{\psi_R}}_\psi + 2m_n^2M_\psi\braket{e^Ag^{(n)}g^{(n)}f^{(m)}_{\psi_L}f^{(m)}_{\psi_R}}_\psi.
    \endaligned
    \end{equation}
    On the other hand, applying eigenequations to the surface integral
    \begin{equation}
        \int_{z_1}^{z_2} dz~\partial_z\left(e^{4A}g^{(n)}g^{(n)}f^{(m)}_{\psi_L}f^{(m)}_{\psi_R}\right) = 0,
    \end{equation}
    one gets
    \begin{equation}
    \aligned
        \braket{A'g^{(n)}g^{(n)}f^{(m)}_{\psi_L}f^{(m)}_{\psi_R}}_\psi =~& - \frac{1}{3}m_n\braket{f^{(n)}g^{(n)}f^{(m)}_{\psi_L}f^{(m)}_{\psi_R}}_\psi +  \frac{1}{6}m_{\psi,m}\braket{g^{(n)}g^{(n)}f^{(m)}_{\psi_R}f^{(m)}_{\psi_R}}_\psi \\
        & -  \frac{1}{6}m_{\psi,m}\braket{g^{(n)}g^{(n)}f^{(m)}_{\psi_L}f^{(m)}_{\psi_L}}_\psi.
    \endaligned
    \end{equation}
    Hence,
    \begin{equation}
    \aligned
        \sum_{j=0}^{\infty}m_{\psi,j}b^{\psi_L\psi_R}_{\bar{n}jm}b^{\psi_L\psi_R}_{\bar{n}mj} =~& - \frac{1}{2}m_n^2m_{\psi,m}\left(\braket{g^{(n)}g^{(n)}f^{(m)}_{\psi_L}f^{(m)}_{\psi_L}}_\psi + \braket{g^{(n)}g^{(n)}f^{(m)}_{\psi_R}f^{(m)}_{\psi_R}}_\psi\right)\\
        &~ + 2m_n^2M_\psi\braket{e^Ag^{(n)}g^{(n)}f^{(m)}_{\psi_L}f^{(m)}_{\psi_R}}_\psi\\
        =~& - \frac{1}{2}m_{\psi,m} \left(b^{\psi_L}_{\bar{n}\bar{n}mm} + b^{\psi_R}_{\bar{n}\bar{n}mm}\right) + 2M_\psi b^{M_\psi}_{\bar{n}\bar{n}mm}.
    \endaligned
    \end{equation}
    Finally, we are ready to prove the sum-rule given in Eq.~(\ref{sr:fermion_5}).
    \begin{equation}
    \aligned
        \sum_{j=0}^{\infty}m_{\psi,j}^3 a^{\psi_L}_{nmj}a^{\psi_R}_{nmj} =~&\sum_{j=0}^{\infty}m_{\psi,j} \left(b^{\psi_L\psi_R}_{\bar{n}mj} + m_{\psi,m}a^{\psi_R}_{nmj}\right)\left(-b^{\psi_L\psi_R}_{\bar{n}jm} + m_{\psi,m}a^{\psi_L}_{nmj}\right) \\
        =~&\sum_{j=0}^{\infty}\left\{-m_{\psi,j}\left(b^{\psi_L\psi_R}_{\bar{n}mj}b^{\psi_L\psi_R}_{\bar{n}jm}\right) +m_{\psi,m}m_{\psi,j}^2\left[\left(a^{\psi_L}_{nmj}\right)^2 + \left(a^{\psi_R}_{nmj}\right)^2\right]\right.\\
        &~ \left. - m_{\psi,m}^2m_{\psi,j}a^{\psi_L}_{nmj} a^{\psi_R}_{nmj}\right\} \\
        =~& \frac{3}{2}m_{\psi,m}\left(b^{\psi_{L}}_{\bar{n}\bar{n}mm} + b^{\psi_{R}}_{\bar{n}\bar{n}mm}\right) + \frac{1}{2}m_{\psi,m}^3\left(a^{\psi_{L}}_{nnmm} + a^{\psi_{R}}_{nnmm}\right)\\
        &~ - 2M_\psi b^{M_\psi}_{\bar{n}\bar{n}mm}.
    \endaligned
    \end{equation}
    \item 
    The proof of the the radion sum-rule given in Eq.~(\ref{sr:fermion_6}).
    
    {\bf Proof:}  Applying eigenequations to the surface integral
    \begin{equation}
         \int_{z_1}^{z_2}dz \,\left[\partial_z\left(e^{3A}f^{(n)}f^{(n)}g^{(j)}\right)\right] = \int_{z_1}^{z_2}dz\, \left[\partial_z\left(e^{3A}g^{(n)}g^{(n)}g^{(j)}\right)\right] = 0~,
     \end{equation}
     one gets
     \begin{equation}
         \braket{A'g^{(n)}g^{(n)}g^{(j)}} = -\dfrac{m_j}{6}\left(\braket{f^{(n)}f^{(n)}f^{(j)}} + \braket{g^{(n)}g^{(n)}f^{(j)}}\right).
     \end{equation}
     Then, from 
     \begin{equation}
         \aligned
         \int_{z_1}^{z_2}dz \,\left[\partial_z\left(e^{4A}g^{(j)}f^{(m)}_{\psi_{L/R}}f^{(m)}_{\psi_{L/R}}\right)\right] &= 0~,\\
         \int_{z_1}^{z_2}dz \,\left[\partial_z\left(e^{5A}g^{(j)}f^{(m)}_{\psi_L}f^{(m)}_{\psi_R}\right)\right] &= 0~,
         \endaligned
     \end{equation}
     one gets
     \begin{equation}
         \aligned
         \braket{A'g^{(j)}f^{(m)}_{\psi_{L/R}}f^{(m)}_{\psi_{L/R}}}_\psi =~& -\dfrac{m_j}{3}\braket{f^{(j)}f^{(m)}_{\psi_{L/R}}f^{(m)}_{\psi_{L/R}}}_\psi \pm \dfrac{2m_{\psi,m}}{3}\braket{g^{(j)}f^{(m)}_{\psi_L}f^{(m)}_{\psi_R}}_\psi\\
         &~ \mp \dfrac{2M_{\psi}}{3}\braket{e^Ag^{(j)}f^{(m)}_{\psi_{L/R}}f^{(m)}_{\psi_{L/R}}}_\psi,
         \endaligned
     \end{equation}
     \begin{equation}
         \aligned
         \braket{e^AA'g^{(j)}f^{(m)}_{\psi_L}f^{(m)}_{\psi_R}}_\psi =~& -\dfrac{m_j}{2}\braket{e^Af^{(j)}f^{(m)}_{\psi_L}f^{(m)}_{\psi_R}}_\psi- \dfrac{m_{\psi,m}}{2}\braket{e^Ag^{(j)}f^{(m)}_{\psi_L}f^{(m)}_{\psi_L}}_\psi\\
         &~ + \dfrac{m_{\psi,m}}{2}\braket{e^Ag^{(j)}f^{(m)}_{\psi_R}f^{(m)}_{\psi_R}}_\psi.
         \endaligned
     \end{equation}
     \sout{Therefore, }
     Note that, by combining the SUSY relations
     \begin{equation}
         \begin{dcases}
             -(\partial_z+3A')g^{(j)} = m_j f^{(j)}, \\
             (\partial_z + A')g^{(j)} = m_j k^{(j)},
         \end{dcases}
         \label{eq:k-fg-relationi}
     \end{equation}
     one gets
     \begin{equation}
         k^{(j)} = -f^{(j)} -\dfrac{2A'}{m_j}g^{(j)}\quad{\rm for~}j> 0.
         \label{eq:k-fg-relation}
     \end{equation}
     Thus, we have
     \begin{equation}
     \aligned
         m_n^2\braket{g^{(n)}g^{(n)}k^{(j)}} =~& - m_n^2\braket{g^{(n)}g^{(n)}f^{(j)}} - \dfrac{2m_n^2}{m_j}\braket{A'g^{(n)}g^{(n)}g^{(j)}}\\
         =~& -\dfrac{2m_n^2}{3}\braket{g^{(n)}g^{(n)}f^{(j)}} + \dfrac{m_n^2}{3}\braket{f^{(n)}f^{(n)}f^{(j)}}\\
         =~&-\frac{2}{3}b_{\bar{n}\bar{n}j} + \frac{m^2_n}{3}a_{nnj}\qquad({\rm for~}j> 0)~.
     \endaligned
     \end{equation}
     And,
     \begin{equation}
     \aligned
         &\braket{k^{(j)}(f^{(m)}_{\psi_L}f^{(m)}_{\psi_L} + f^{(m)}_{\psi_R}f^{(m)}_{\psi_R}-\frac{4M_\psi}{3m_{\psi,m}}e^Af^{(m)}_{\psi_L}f^{(m)}_{\psi_R})} \vphantom{braket{f^{(m)}_{\psi_L}}}_\psi \\
         =~& - \braket{f^{(j)}(f^{(m)}_{\psi_L}f^{(m)}_{\psi_L} + f^{(m)}_{\psi_R}f^{(m)}_{\psi_R}-\frac{4M_\psi}{3m_{\psi,m}}e^Af^{(m)}_{\psi_L}f^{(m)}_{\psi_R})}\vphantom{braket{f^{(m)}_{\psi_L}}}_\psi \\
         &- \dfrac{2}{m_j}\braket{A'g^{(j)}(f^{(m)}_{\psi_L}f^{(m)}_{\psi_L} + f^{(m)}_{\psi_R}f^{(m)}_{\psi_R}-\frac{4M_\psi}{3m_{\psi,m}}e^Af^{(m)}_{\psi_L}f^{(m)}_{\psi_R})}\vphantom{braket{f^{(m)}_{\psi_L}}}_\psi\\
         =~& -\dfrac{1}{3}\braket{f^{(j)}(f^{(m)}_{\psi_L}f^{(m)}_{\psi_L} + f^{(m)}_{\psi_R}f^{(m)}_{\psi_R})}_\psi \\
         =~&-\frac{1}{3} \left(a^{\psi_L}_{jmm} + a^{\psi_R}_{jmm}\right)\qquad({\rm for~}j> 0)~.
     \endaligned
     \end{equation}
     Finally, using the completeness of the wavefucntions $k^{(j)}$ of the scalar Golstone boson, we have
     \begin{equation}
     \aligned
       & b_{\bar{n}\bar{n}r}\left(a^{\psi_L}_{mmr} + a^{\psi_R}_{mmr}-\frac{4M_\psi}{3m_{\psi,m}}a^{M_\psi}_{mmr}\right)\\
       =~& m_n^2\braket{g^{(n)}g^{(n)}k^{(0)}}\braket{k^{(0)}(f^{(m)}_{\psi_L}f^{(m)}_{\psi_L} + f^{(m)}_{\psi_R}f^{(m)}_{\psi_R}-\frac{4M_\psi}{3m_{\psi,m}}e^Af^{(m)}_{\psi_L}f^{(m)}_{\psi_R})}\vphantom{braket{f^{(m)}_{\psi_L}}}_\psi\\
       =~& m_n^2\sum_{j=0}^{\infty}\braket{g^{(n)}g^{(n)}k^{(j)}}\braket{k^{(j)}(f^{(m)}_{\psi_L}f^{(m)}_{\psi_L} + f^{(m)}_{\psi_R}f^{(m)}_{\psi_R}-\frac{4M_\psi}{3m_{\psi,m}}e^Af^{(m)}_{\psi_L}f^{(m)}_{\psi_R})}\vphantom{braket{f^{(m)}_{\psi_L}}}_\psi \\
       &~ - m_n^2\sum_{j=1}^{\infty}\braket{g^{(n)}g^{(n)}k^{(j)}}\braket{k^{(j)}(f^{(m)}_{\psi_L}f^{(m)}_{\psi_L} + f^{(m)}_{\psi_R}f^{(m)}_{\psi_R}-\frac{4M_\psi}{3m_{\psi,m}}e^Af^{(m)}_{\psi_L}f^{(m)}_{\psi_R})}\vphantom{braket{f^{(m)}_{\psi_L}}}_\psi\\
       =~& m_n^2\braket{g^{(n)}g^{(n)}(f^{(m)}_{\psi_L}f^{(m)}_{\psi_L} + f^{(m)}_{\psi_R}f^{(m)}_{\psi_R}-\frac{4M_\psi}{3m_{\psi,m}}e^Af^{(m)}_{\psi_L}f^{(m)}_{\psi_R})}\vphantom{braket{f^{(m)}_{\psi_L}}}_\psi \\
       &~~  + \frac{1}{9}\sum_{j=1}^{\infty}\left(m_n^2a_{nnj} - 2 b_{\bar{n}\bar{n}j}\right)\left(a^{\psi_L}_{jmm} + a^{\psi_R}_{jmm}\right) \\
       =~& \left(b^{\psi_L}_{\bar{n}\bar{n}mm} + b^{\psi_R}_{\bar{n}\bar{n}mm}-\frac{4M_\psi}{3m_{\psi,m}}b^{M_\psi}_{\bar{n}\bar{n}mm}\right) + \frac{1}{9}\sum_{j=0}^{\infty}\left(m_n^2a_{nnj} - 2 b_{\bar{n}\bar{n}j}\right)\left(a^{\psi_L}_{jmm} + a^{\psi_R}_{jmm}\right) \\
       &~~ - \frac{1}{9}\left(m_n^2a_{nn0} - 2 b_{\bar{n}\bar{n}0}\right)\left(a^{\psi_L}_{0mm} + a^{\psi_R}_{0mm}\right) \\
       =~& \frac{7}{9}\left(b^{\psi_L}_{\bar{n}\bar{n}mm} + b^{\psi_R}_{\bar{n}\bar{n}mm}\right) + \frac{1}{9}m_n^2a_{nn0}\left(a^{\psi_L}_{0mm}+a^{\psi_R}_{0mm}\right)\\
        &~~ + \frac{1}{9}m_n^2\left(a^{\psi_L}_{nnmm}+a^{\psi_R}_{nnmm}\right)-\frac{4M_\psi}{3m_{\psi,m}}b^{M_\psi}_{\bar{n}\bar{n}mm}.
     \endaligned
     \end{equation}

     \item 
     In a similar manner, we can also prove the the radion sum-rule for brane matter given in Eq.~(\ref{sr:brane_3}).

     \begin{equation}
    \aligned
    b_{\bar{n}\bar{n}r}k^{(0)}(\bar{z}) =~& m_n^2\braket{g^{(n)}g^{(n)}k^{(0)}} k^{(0)}(\bar{z})\\
    =~& \sum_{j=0}^{\infty}m_n^2\braket{g^{(n)}g^{(n)}k^{(j)}} k^{(j)}(\bar{z}) - \sum_{j=1}^{\infty}m_n^2\braket{g^{(n)}g^{(n)}k^{(j)}} k^{(j)}(\bar{z}) \\
    =~& m_n^2\left[g^{(n)}(\bar{z})\right]^2 - \sum_{j=1}^{\infty}\left(-\frac{2}{3}b_{\bar{n}\bar{n}j} + \frac{m^2_n}{3}a_{nnj}\right) \left(-f^{(j)}(\bar{z})\right) \\
    =~&\sum_{j=0}^{\infty}\left(-\frac{2}{3}b_{\bar{n}\bar{n}j} + \frac{m^2_n}{3}a_{nnj}\right) f^{(j)}(\bar{z})+ \left(-\frac{2}{3}b_{\bar{n}\bar{n}0} - \frac{m^2_n}{3}a_{nn0}\right) f^{(0)}(\bar{z}) \\
    =~& \frac{m_n^2}{3}\left[f^{(n)}(\bar{z})\right]^2 + \frac{m_n^2}{3}a_{nn0}f^{(0)}(\bar{z}).
    \endaligned
    \end{equation}

\end{itemize}

\newpage

\bibliographystyle{apsrev4-1.bst}

\bibliography{main}{}

\end{document}